\documentclass[aps,pra,twocolumn,amsmath,showpacs,showkeys,superscriptaddress,longbibliography,floatfix]{revtex4-1}

\usepackage{amsmath}
\usepackage{amssymb}
\usepackage{physics}
\usepackage{hyperref}
\usepackage{cleveref}

\usepackage{graphicx}

\usepackage[english]{babel}

\usepackage{color}
\usepackage[squaren]{SIunits}

\crefname{section}{Sec.}{Sections}
\crefname{equation}{Eq.}{Eqs.}
\Crefname{equation}{Equation}{Equations}
\crefname{figure}{Fig.}{Figs.}
\Crefname{figure}{Figure}{Figures}

\begin{document}

\newcommand{\highindex}[2]{#1^{(#2)}}
\newcommand{\te}[1]{\text{#1}}
\newcommand{\kb}[2]{\ket{#1}\bra{#2}}
\newcommand{\kom}[2]{\left[ #1, #2 \right]}
\newcommand{\erw}[1]{\langle {#1} \rangle}
\newcommand{\korr}[1]{\delta \erw{#1}}
\newcommand{\wigner}[3]{{#1}^{(#2)}\left( #3 \right)}
\newcommand{\coeff}[3]{#1_{#2}^{(#3)}}
\newcommand{\ccoeff}[3]{#1_{#2}^{*(#3)}}

\title{Comparison of the semiclassical and quantum optical field dynamics in a pulse-excited optical cavity with a finite number of quantum emitters}

\author{K.~J\"urgens}
\author{F.~Lengers}
\author{D.~Groll}
\author{D.~E.~Reiter}
\affiliation{Institut f\"ur Festk\"orpertheorie, Universit\"at M\"unster,
Wilhelm-Klemm-Str.~10, 48149 M\"unster, Germany}
\author{D.~Wigger}
\affiliation{Department of Theoretical Physics, Wroc\l{}aw University of Science and Technology, 
Wybrze\.{z}e Wyspia\'{n}skiego 27, 50-370 Wroc\l{}aw, Poland}
\author{T.~Kuhn}
\affiliation{Institut f\"ur Festk\"orpertheorie, Universit\"at M\"unster,
Wilhelm-Klemm-Str.~10, 48149 M\"unster, Germany}

\begin{abstract}
The spectral and temporal response of a set of $N$ quantum emitters embedded in a photonic cavity is studied.
Quantum mechanically, such systems can be described by the Tavis-Cummings (TC) model of $N$ two-level systems coupled to a single light mode.
Here we compare the full quantum solution of the TC model for different numbers of quantum emitters with its semiclassical limit after a pulsed excitation of the cavity mode.
Considering different pulse amplitudes, we find that the spectra obtained from the TC model approach the semiclassical one for an increasing number of emitters $N$.
Furthermore they match very well for small pulse amplitudes.
While we observe a very good agreement in the temporal dynamics for photon numbers much smaller than $N$, considerable deviations occur in the regime of photon numbers similar to or larger than $N$, which are linked to collapse and revival phenomena.
Wigner functions of the light mode are calculated for different scenarios to analyze the quantum state of the light field.
We find strong deviations from a coherent state even if the dynamics of the expectation values are still well described by the semiclassical limit.
For higher pulse amplitudes Wigner functions similar to those of Schr\"odinger cat states between two or more quasi-coherent contributions build up. 
\end{abstract}

\maketitle

\section{Introduction}
The interaction of a single light mode with a single or multiple quantum emitters (QEs) is a fundamental model that is successfully applied to numerous systems.
Semiconductor quantum dots (QDs) in a microcavity are a prominent realization of such quantum light sources with a variety of different applications ranging from QD lasers \cite{chow2014emi,czerniuk2017pic} involving QDs in laterally extended cavities formed by Bragg mirrors down to single QE structures for applications in the field of quantum information technology \cite{michler2017qua}.
Color centers of different types placed in a microcavity, such as nitrogen-vacancy \cite{englund2010det,beha2012dia,fehler2019eff,janitz2020cav} or silicon-vacancy centers \cite{lee2012opt, fehler2020pur} in diamond, defect states in hexagonal boron nitride \cite{proscia2020mic,froch2020cou}, or organic molecules \cite{pino2018ten} constitute another class of QEs coupled to a single light mode, which are currently extensively studied.
Increasing possibilities to create deterministic photon emitter structures, e.g., by strain-patterning of $2$D semiconductors \cite{branny2017det,palacios2017lar,kern2016nan}, by lithographic positioning of nanodiamonds \cite{schrinner2020int} or by irradiation with a narrow helium beam \cite{klein2019sit} also pave the way for studying systems with well-defined finite numbers of photon emitters.
Various types of cavities have been realized to host QEs, such as photonic crystal cavities or micropillar cavities \cite{lodahl2015int}.

While structures with many emitters are often treated in terms of a semiclassical model, in which the emitters are described by quantum mechanical few-level systems interacting with a classical electromagnetic field \cite{gehrig2002mes,cartar2017sel,jurgens2020sem}, a single emitter in a cavity is usually described in terms of quantum optics by the Jaynes-Cummings (JC) model \cite{jaynes1963com,shore1993the,groll2020fou}.
Such a quantum optical treatment is necessary because of the strong correlations between the quantum states of the emitter and the photon field inside the cavity.
Between these limiting cases there is an interesting class of systems represented by a finite number of emitters $N$ in a cavity.
Based on the experience in the limiting cases, one would expect that for small $N$ a quantum optical treatment is necessary while for increasing $N$ the dynamical behavior approaches the one obtained from a semiclassical treatment.
In this paper we address this question by analyzing the optically induced dynamics of systems with $1$ up to $60$ QEs in a cavity on a quantum optical level and comparing the results with a semiclassical treatment.
We demonstrate that the spectra obtained by the quantum optical model approach the semiclassical result for increasing number of emitters, however for the numbers studied here they are still much broader.
Furthermore, we show that quantum effects play a crucial role for the photon states as revealed by a Wigner function analysis.

In a previous study we have performed semiclassical simulations of a planar ensemble of QEs, modeled as two-level systems (TLSs), in a one-dimensional photonic cavity driven by a short external laser pulse coupled in through one of the mirrors \cite{jurgens2020sem}.
We have shown that there exists a sharp transition in this semiclassical undamped model, associated with characteristic changes in the spectrum of the light field when varying the pulse amplitudes.
For small amplitudes exciton-polariton-like spectra occur, while Rabi oscillations emerge for large amplitudes.
We have furthermore shown that a sharp transition remains also for an ensemble of QEs with a Gaussian energy distribution.

In the semiclassical limit the dynamics depend on the density of QEs, represented by a continuous number, which is related to the fact that also the light field is a continuous classical variable.
In a quantum optical treatment the light field is described in terms of Fock states involving discrete numbers of photons, which can be absorbed or emitted by the QEs, and the number $N$ of QEs becomes a relevant quantity.
Because of the tensor product structure of the Hilbert space of composite quantum systems, the complexity increases exponentially with increasing $N$, which provides strong limitations on the analytical and numerical feasibility.
Therefore, additional restrictions of the model are necessary, which can be: (i) a limitation to very small numbers $N$ \cite{youssef2010som,albert2013mic,droenner2017col}, (ii) a limitation to small excitation intensities (i.e., small numbers of photons compared to the number of QEs) \cite{tsyplyatyev2009dyn,tsyplyatyev2010cla},or (iii) assuming ensembles consisting of $N$ QEs with identical \cite{tavis1968exa} or a small number of distinct \cite{dhar2018var} frequencies.
Here we will rely on the assumption of an ensemble with identical frequencies, which is further motivated by our finding in Ref.~\cite{jurgens2020sem} that the general features observed when moving from the low-excitation to the high-excitation regime are preserved also for a QE ensemble of non-vanishing width.

In this work we compare the spectra and the dynamics of the semiclassical and quantum optical solutions of $N$ QEs coupled to a single mode of the electric field after a pulsed excitation of the cavity.
The laser pulse creates a coherent field inside the cavity \cite{groll2020fou}, considered as an initial condition for the electric field, which then initiates the dynamics in the coupled QE-light system.

The paper is organized as follows.
In Sec.~\ref{sec.model} we introduce the model and derive the equations of motion, showing that the semiclassical limit is obtained as the mean-field approximation of the full equations of motion.
Then the exact solution of the full quantum system with $N$ QEs, i.e., the Tavis-Cummings (TC) model, is briefly reviewed and applied to the present initial conditions.
Section~\ref{sec.comparison} provides a detailed comparison of the spectral and temporal characteristics in the semiclassical and the quantum optical model for QE numbers between $N=1$ and $60$.
In Sec.~\ref{sec.wigner} the Wigner function is introduced and used to obtain information on the full quantum state of the light field during the initial dynamics.
Section~\ref{sec.collapse} extends this discussion to later times by analyzing the quantum state during the collapse and revival of the field amplitude. While up to this section we concentrate on a Hamiltonian model, in Sec.~\ref{sec.dissipation} we extend the model by including dissipation.
Finally, in Sec.~\ref{sec.conclusions} we briefly summarize our results.

\section{The Tavis-Cummings Model}\label{sec.model}
We consider an ensemble of $N$ QEs treated as TLSs with transition energies $\hbar \highindex{\omega_x}{k}$, coupled with strengths $\hbar \highindex{g}{k}$ to a single light field mode with energy $\hbar \omega_0$ and annihilation (creation) operator $\hat{a}$ $(\hat{a}^{\dagger})$.
This leads to the Hamiltonian of the TC model \cite{tavis1968exa,krimer2019cri,zens2019cri}
\begin{align}
 \hat{H} &= \hbar \omega_0 \hat{a}^\dagger \hat{a} + \frac{\hbar}{2} \sum_{k=1}^N \highindex{\omega_x}{k} \highindex{\hat{\sigma}_3}{k} \nonumber \\
 &\quad + \hbar \sum_{k=1}^N \highindex{g}{k} \left( \hat{a} \highindex{\hat{\sigma}_+}{k} + \hat{a}^\dagger \highindex{\hat{\sigma}_-}{k} \right), \label{eq:H_ens}
\end{align}
where $\highindex{\hat{\sigma}_\pm}{k}=\frac{1}{2}\left(\highindex{\hat{\sigma}_1}{k}\pm i \highindex{\hat{\sigma}_2}{k}\right)$, and $\{\highindex{\hat{\sigma}_j}{k}, j=1,2,3\}$ denote the Pauli matrices acting on QE $k$.
They are directly related to the inversion ($\highindex{\hat{\sigma}_3}{k}$) and to the polarization ($\highindex{\hat{\sigma}_\pm}{k}$) of the QEs.
In general, the transition frequencies and coupling strengths may be different for each QE \cite{zens2019cri,krimer2019cri,eastham2009qua,dhar2018var}.
Note that the Hamiltonian in \cref{eq:H_ens} involves the standard rotating wave approximation (RWA); the Hamiltonian for the QE-light coupling without RWA is called Dicke Hamiltonian \cite{dicke1954coh,bastarrachea2014com1,bastarrachea2014com2}.
Depending on the system parameters, the counter rotating terms can have an influence on the ground state of the system \cite{bastarrachea2014com1,bastarrachea2014com2} or its dynamics \cite{seke1995squ,seke1997eff,crescente2020ult}.
For the parameters studied here, however, the RWA can be assumed to be well justified, as is also confirmed by our calculations within the semiclassical model \cite{jurgens2020sem}.

\begin{figure}
 \centering
 \includegraphics[width=1\columnwidth]{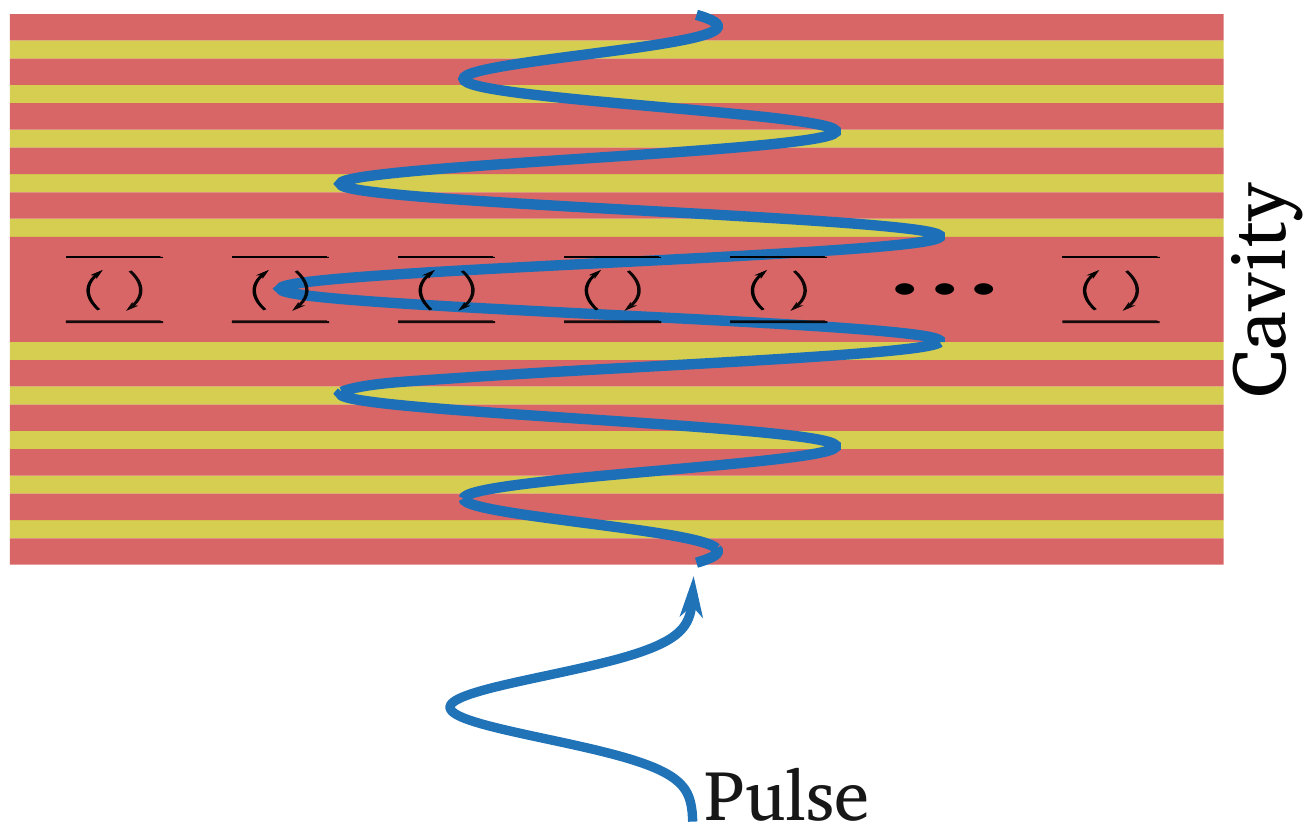}
 \caption{Sketch of the system. $N$ QEs are coupled to a single light mode inside a cavity. The cavity field is excited using a short external laser pulse.}
 \label{fig1}
\end{figure}

The system is excited using a short laser pulse coupled in through one of the mirrors, such that the cavity mode is initialized in a coherent state \cite{groll2020fou}, as schematically shown in \cref{fig1}.
The generated light field then couples to the QEs and starts the internal dynamics.
As has been seen in the semiclassical case in Ref.~\cite{jurgens2020sem}, for a pulse which is shorter than the characteristic time scale of the dynamics, the excitation can be very well approximated by taking the excited cavity mode as an initial condition and disregarding the details of the injection process.
We will follow this scheme also here.
Therefore, the driving laser field is not included in the Hamiltonian given in \cref{eq:H_ens}.
This is different from studies investigating the case of a constant driving \cite{zens2019cri,krimer2019cri}.
The influence of initial Fock- and coherent states are discussed in Ref.~\cite{seke1989col}, where collapse and revival phenomena of different origin are found.
Other excitation schemes like chirped excitation of the QEs \cite{eastham2009qua} or different initial conditions \cite{keeling2009qua} have also been studied.

\subsection{Equations of motion and semiclassical limit}
To clearly see the relation between the quantum optical and the semiclassical model we start by deriving the equations of motion for the expectation values of the operators $\hat{a}$, $\highindex{\hat{\sigma}_3}{k}$ and $\highindex{\hat{\sigma}_-}{k}$, i.e., the coherent field amplitude, as well as the inversion and the polarization of QE $k$, respectively.
Using Ehrenfest's theorem in the rotating frame with $\hat{a} = \hat{\tilde{a}} e^{-i\omega_0 t}$, $\highindex{\hat{\sigma}_-}{k} = \highindex{\hat{\tilde{\sigma}}_-}{k} e^{-i\omega_0 t}$, the equations of motion read
\begin{subequations} 
\begin{align}
    \dv{t} \erw{\hat{\tilde{a}}} &= -i \sum_{k=1}^N \highindex{g}{k} \erw{\highindex{\hat{\tilde{\sigma}}_-}{k}} , \\
    \dv{t} \erw{\highindex{\hat{\sigma}_3}{k}} &= 2 i \highindex{g}{k} \left[ \erw{\hat{\tilde{a}}^\dagger \highindex{\hat{\tilde{\sigma}}_-}{k}} - \erw{\hat{\tilde{a}} \highindex{\hat{\tilde{\sigma}}_+}{k}} \right] , \\
    \dv{t} \erw{\highindex{\hat{\tilde{\sigma}}_-}{k}} &= -i \left(\highindex{\omega_x}{k}-\omega_0\right) \erw{\highindex{\hat{\tilde{\sigma}}_-}{k}} + i \highindex{g}{k} \erw{\hat{\tilde{a}} \highindex{\hat{\sigma}_3}{k}}.
    \end{align}
\end{subequations}
Obviously, this set of equations of motion is not closed, instead on the right hand side it involves expectation values of products of a field and a QE operator.
These terms describe the influence of quantum correlations between light and matter on the dynamics.
By separating these quantities via $\erw{\hat{x} \hat{y}} = \erw{\hat{x}} \erw{\hat{y}} + \korr{\hat{x} \hat{y}}$ into mean field contributions and correlations, where $\hat{x}$ and $\hat{y}$ denote any two operators, and neglecting the correlations we arrive at
\begin{subequations} 
\begin{align}
    \dv{t} \erw{\hat{\tilde{a}}} &= -i \sum_{k=1}^N \highindex{g}{k} \erw{\highindex{\hat{\tilde{\sigma}}_-}{k}} , \\
    \dv{t} \erw{\highindex{\hat{\sigma}_3}{k}} &= 2 i \highindex{g}{k} \left[ \erw{\hat{\tilde{a}}^\dagger} \erw{\highindex{\hat{\tilde{\sigma}}_-}{k}} - \erw{\hat{\tilde{a}}} \erw{\highindex{\hat{\tilde{\sigma}}_+}{k}} \right] , \\
    \dv{t} \erw{\highindex{\hat{\tilde{\sigma}}_-}{k}} &= -i \left(\highindex{\omega_x}{k}-\omega_0\right) \erw{\highindex{\hat{\tilde{\sigma}}_-}{k}} + i \highindex{g}{k} \erw{\hat{\tilde{a}}} \erw{\highindex{\hat{\sigma}_3}{k}}.
    \end{align}
\end{subequations}
The dynamics of $\erw{\hat{\tilde{a}}}$, $\erw{\highindex{\hat{\sigma}_3}{k}}$ and $\erw{\highindex{\hat{\tilde{\sigma}}_-}{k}}$ only depend on the exciton frequency of the QE $\highindex{\omega_x}{k}$ and its coupling constant $\highindex{g}{k}$.
Assuming that $\highindex{g}{k}$ only depends on its frequency $\highindex{\omega_x}{k}$, i.e., $\highindex{g}{k} = g\left(\highindex{\omega_x}{k}\right)$, it is possible to characterize the system using the continuous variable $\omega_x$ instead of the index $k$ of each QE assigning $\erw{\hat{\sigma}_3} (\omega_x) = \erw{\highindex{\hat{\sigma}_3}{k}}$ and $\erw{\hat{\sigma}_\pm} (\omega_x) = \erw{\highindex{\hat{\sigma}_\pm}{k}}$ for $\omega_x = \highindex{\omega_x}{k}$.

Now we rescale the dynamical variables and the time, according to
\begin{subequations} \label{eq:transformation}
\begin{align}
 \alpha &= \frac{1}{\sqrt{N}} \erw{\hat{\tilde{a}}}, \\
 \beta(\omega_x) &= \erw{\hat{\sigma}_3}(\omega_x), \\
 \gamma(\omega_x) &= -i \erw{\hat{\tilde{\sigma}}_-} (\omega_x), \\
 \tau &= G \sqrt{N} t, \label{eq:trafo_tau}
\end{align}
\label{eq:trafo}%
\end{subequations}
where $G = \frac{1}{N} \sum_{k=1}^N \highindex{g}{k}$ is the mean coupling strength.
The variable $\alpha$, which is the photon amplitude normalized to the square root of $N$, will be important throughout the paper.
$\tau$ denotes the scaled time.
This leads to the equations of motion
\begin{subequations}
    \begin{align}
    \dv{\tau} \alpha(\tau) &= \int \dd{\omega_x} \rho_{\rm QE}(\omega_x) \gamma(\omega_x,\tau) ,\\
    \dv{\tau} \beta(\omega_x,\tau) &= -2\frac{g(\omega_x)}{G} \left[ \alpha^*(\tau) \gamma(\omega_x,\tau) + \alpha(\tau)  \gamma^*(\omega_x,\tau) \right] ,\\
    \dv{\tau} \gamma(\omega_x,\tau) &= -i \frac{\omega_x-\omega_0}{G\sqrt{N}} \gamma(\omega_x,\tau) + \frac{g(\omega_x)}{G} \alpha(\tau) \beta(\omega_x,\tau),
    \end{align}
    \label{eq:eom_semiclassical}%
\end{subequations}
where $\rho_{\rm QE} = \frac{1}{GN} \sum_{k=1}^N \highindex{g}{k} \delta(\omega_x - \highindex{\omega_x}{k})$ is the effective spectral QE distribution.
Comparing this result with the equations obtained from a semiclassical model for a QE ensemble in a Bragg mirror cavity [Eqs.~(34) in Ref.~\cite{jurgens2020sem}], we find that they are in complete agreement when setting $G = M \sqrt{\frac{u_c(z_0) v_c^*(z_0) \omega_0}{2\hbar A n(z_0)^2 \epsilon_0}}$, where $M$ is the mean value of the dipole moments, $A$ is a normalization area, $n(z_0)$ is the refractive index at the position of the QEs, $u_c(z_0)$ is the eigenmode of the wave equation at the position of the QE layer $z_0$ and $v_c(z_0)=-\frac{d^2}{dz^2}u_c(z_0)$ is the eigenmode of the adjoint wave equation.
Note that in Ref.~\cite{jurgens2020sem} furthermore a frequency-independent coupling matrix element  $g(\omega_x)=G$ has been assumed.

Like in the semiclassical case studied in Ref.~\cite{jurgens2020sem}, here we are interested in the dynamics of the system after exciting the cavity with a short laser pulse.
In practice, we will assume that the pulse creates a coherent state inside the cavity, such that the system can initially be described by the coherent amplitude $\xi = \erw{a}(0) = \alpha_0\sqrt{N}$ of the field \cite{groll2020fou}, where $\left|\alpha_0\right|^2$ is the mean initial photon number per QE, with all QEs being in their respective ground state.
This corresponds to the initial conditions $\alpha(\tau=0)=\alpha_0$, $\beta(\omega_x,\tau=0)=-1$ and $\gamma(\omega_x,\tau=0)=0$.
Pulsed excitations, where instead of the cavity an ensemble of QEs is directly pumped, have been studied, e.g., in Ref.~\cite{eastham2009qua}.

The semiclassical (or mean field) equations shown in Eqs.~(\ref{eq:eom_semiclassical}) can be solved numerically for arbitrary ensemble shapes $\rho_{\rm QE}(\omega_x) $ using standard Runge-Kutta methods \cite{jurgens2020sem}.

In contrast to the semiclassical case, a complete quantum optical solution of the Hamiltonian $\hat{H}$ of \cref{eq:H_ens} for arbitrary spectral distributions and numbers $N$ of QEs is difficult because of the exponential increase of Hilbert space dimensionality with increasing $N$.
The quantum system with arbitrary frequency distributions has been investigated in Refs.~\cite{tsyplyatyev2009dyn,tsyplyatyev2010cla} where, however, only weak excitations, i.e., small numbers of photons compared to the number of QEs, have been studied.
Here we are interested in the comparison between semiclassical and quantum optical behavior from low excitations up to the Rabi oscillation regime, the latter being characterized by photon numbers larger than $N$.
Motivated by the findings in Ref.~\cite{jurgens2020sem} that the general features of the spectra are preserved also in the presence of spectral broadening, we restrict our study to the case of identical QEs with $\highindex{\omega_x}{k} = \omega_x$, $\highindex{g}{k} = g=G$ and $\omega_x = \omega_0$ being resonant to the cavity mode.
Then the model reduces to the standard TC model, for which a full solution is available.

The TC and Dicke models include a normal- to super-radiant phase-transition in the semiclassical limit ($N \rightarrow \infty$) \cite{bastarrachea2014com1,bastarrachea2014com2,castanos2009coh,emary2003qua,zou2013qua}.
This phase transition is hard to realize experimentally, therefore also coherently driven systems in the presence of damping have been studied \cite{krimer2019cri,zens2019cri,munoz2019sym} and phase transitions and hysteresis effects were found depending on the amplitude of the driving field \cite{krimer2019cri,zens2019cri,zou2013qua}.
It was shown that quantum effects are negligible outside a critical region of the coherent amplitude, when the number of two-level systems $N$ is large \cite{zens2019cri}.
Far from this critical order parameter the system can then be described by its semiclassical limit $N \rightarrow \infty$.
In contrast to these equilibrium phase transitions or phase transitions between stationary non-equilibrium states, the transition discussed in Ref.~\cite{jurgens2020sem} occurs between different types of time dependent states.

Beside these ground state and steady-state behaviors, Dicke and TC models also show interesting dynamical quantum features like squeezing \cite{retamal1997sqe,hassan1993per,genes2003spi,seke1995squ,seke1997eff,ramon1998col}, collapse and revival phenomena \cite{agarwal2012tav,ramon1998col,meunier2006ent,jarvis2009dyn} or superradiance and subradiance \cite{temnov2005sup,gegg2018sup}.
We will come back to some of these features below.

\subsection{Derivation of the quantized model}
For identical QEs the Hamiltonian of \cref{eq:H_ens} agrees with the standard TC model \cite{tavis1968exa} and can be written in an angular momentum representation 
\begin{align}
 \hat{H} = \hbar \omega_0 \hat{a}^\dagger \hat{a} + \hbar \omega_x \hat{J}_3 + \hbar g \left( \hat{a} \hat{J}_+ + \hat{a}^\dagger \hat{J}_- \right) \label{eq:H} 
\end{align}
with total angular momentum operators 
\begin{align}
 \hat{J}_3 = \frac{1}{2} \sum_{k=1}^N \highindex{\hat{\sigma}_3}{k},  &&
 \hat{J}_\pm = \sum_{k=1}^N \highindex{\hat{\sigma}_\pm}{k}. \nonumber 
\end{align}
The total Hilbert space $\mathcal{H}$ is a tensor product of the Hilbert space of the photons $\mathcal{H}_\te{Phot}$ and the Hilbert space of the QEs $\mathcal{H}_\te{QE}$.
The Fock states with $n$ photons $\ket{n}$ constitute a basis of $\mathcal{H}_\te{Phot}$, whereas a basis of $\mathcal{H}_\te{QE}$ is given by the angular momentum states $\ket{j, m}$ characterized by the total angular momentum quantum number $j$ and the eigenvalue $m$ of the third component $\hat{J}_3$.
The Hamiltonian preserves the quantum number $j$ and, assuming all QEs to be in the ground state at $t=0$, the quantum number $j$ is given by half of the total number of QEs $j=N/2$.
The quantum number $m= -j$ then corresponds to the state with zero excitations and $m=j$ corresponds to the case of fully inverted QEs.
In the basis of the product states $\ket{n}\ket{j, m}$ the Hamiltonian is tridiagonal and the subspaces with constant $n + m + j$ decouple.
Here we classify the subspaces by the maximum number of photons $n$, i.e., the number of photons for the case that all QEs are in their respective ground state, implying $m=-j$.
These subspaces can be diagonalized either analytically \cite{tavis1968exa} or numerically leading to the eigenbasis $\ket{j, n; l}$ with
\begin{align}
 \ket{j, n; l} &= \sum_{k=0}^{\min(N, n)} c_{k, l}^{(N,n)} \ket{n-k}\ket{j, -j+k} , \\
 \hat{H}\ket{j, n; l} &= E_l^{(N,n)} \ket{j, n; l}.
\end{align}
In each subspace there are $\min(N, n)+1$ states and the state vector $\ket{\highindex{\Psi}{N}(t)} \in \mathcal{H}$ of the full system has the time dependence
\begin{align}
 \ket{\highindex{\Psi}{N}(t)} = e^{-\frac{i}{\hbar}\hat{H}t} \ket{\highindex{\Psi}{N}(0)}. \label{eq:time_state}
\end{align}
As explained above (and confirmed in the semiclassical model \cite{jurgens2020sem}) we assume that the exciting pulse has no direct influence on the QEs inside the cavity but instantaneously creates a coherent photon state $\ket{\xi} = e^{-\frac{\abs{\xi}^2}{2}} \sum_n \frac{\xi^n}{\sqrt{n!}} \ket{n}$ at time $t=0$ \cite{groll2020fou}.
The coherent amplitude $\xi = \erw{a}(0)$ is the initial condition for the strength of the light field inside the cavity and $\abs{\xi}^2$ corresponds to the initial mean number of photons in the system.
The initial state of the full system is therefore given by the product of the state $\ket{j, -j}$ reflecting the fact that all QEs are in their respective ground state and a coherent state $\ket{\xi}$ for the light field, i.e., $\ket{\highindex{\Psi}{N}(0)} = \ket{\xi} \ket{j, -j}$.
The subsequent time evolution of the state is then given by
\begin{align}
 \ket{\highindex{\Psi}{N}(t)} = e^{-\frac{\abs{\xi}^2}{2}} \sum_{n, l} \frac{\xi^n}{\sqrt{n!}} c_{0, l}^{*(N,n)} e^{-\frac{i}{\hbar}\highindex{E_l}{N,n}t}\ket{j, n; l}. \label{eq:time_evolution_state}
\end{align}
Using the transformation from \cref{eq:transformation}, the expectation value $\highindex{\alpha}{N}(t) = \erw{\hat{\alpha}} = \erw{\hat{\tilde{a}}} / \sqrt{N}$ in the rotating frame reads
\begin{align}
 \highindex{\alpha}{N}(t) &= \frac{e^{-\abs{\xi}^2}}{\sqrt{N}} \sum_{n, l, l'} 
        \frac{\xi^n}{\sqrt{n!}} \frac{\xi^{*(n-1)}}{\sqrt{(n-1)!}}\ccoeff{c}{0, l}{N,n} \coeff{c}{0, l'}{N,n-1} \nonumber \\
        & \times  e^{-\frac{i}{\hbar} \left( E_l^{(N,n)} - E_{l'}^{(N,n-1)}  - \hbar \omega_0 \right)t} B_{l,l'}^{(N, n)}, \label{eq:erw_x_N}
\end{align}
where
\begin{align}
        B_{l,l'}^{(N, n)} &= \sum_{k=0}^{\min(N, n-1)} \ccoeff{c}{k, l'}{N,n-1} \coeff{c}{k, l}{N,n} \sqrt{n-k}. \nonumber
\end{align}
The semiclassical limit, calculated using \cref{eq:eom_semiclassical} with $\rho(\omega_x) = \delta(\omega_x)$, corresponds to the limit $N \rightarrow \infty$ (similar to Ref.~\cite{zens2019cri}) and will therefore be denoted by $\highindex{\alpha}{\infty}(t)$.

For small values $\xi \ll 1$ only the Fock states with $n = 0$ and $n = 1$ contribute to \cref{eq:erw_x_N} leading to
\begin{align}
 \highindex{\alpha}{N}(t) &\approx \xi \frac{e^{-\abs{\xi}^2}}{\sqrt{N}} \sum_{l=0}^{1} \sum_{l'=0}^0  \ccoeff{c}{0,l}{N,1} \coeff{c}{0,l'}{N,0} \nonumber \\
 &\times e^{-\frac{i}{\hbar} \left( E_l^{(N,1)} - E_{l'}^{(N,0)} - \hbar \omega_0 \right) t} B_{l,l'}^{(N, 1)}.
\end{align}
The subspaces with $n=0$ and $n=1$ photons can be diagonalized analytically similar to the JC model \cite{haroche2006exp}, leading to two frequency contributions at 
\begin{align}
\frac{1}{\hbar}\left(E_l^{(N,1)} - E_{0}^{(N,0)} - \hbar \omega_0 \right) = (-1)^l g \sqrt{N} \label{eq:splitting}
\end{align}
for $l \in \{0, 1\}$ whose difference scales with $\sqrt{N}$.
These findings are in good agreement with Refs.~\cite{cummings1983exa,tsyplyatyev2009dyn,tsyplyatyev2010cla}, because in these cases the number of excited QEs is small compared to their total number.
It also fits very well to the semiclassical Dicke model \cite{bakemeier2013dyn} in the limit of weak couplings where the RWA holds.
The limit of small initial values $\highindex{\alpha}{N}(0) = \alpha_0 = \frac{\xi}{\sqrt{N}} \ll 1$ corresponds to the formation of an exciton-polariton.
The splitting depends on the number of QEs, identical to the semiclassical case \cite{jurgens2020sem}.

\section{Comparison of the spectra and the dynamics}\label{sec.comparison}

\subsection{Spectra}
\begin{figure}
 \centering
 \includegraphics[width=1\columnwidth]{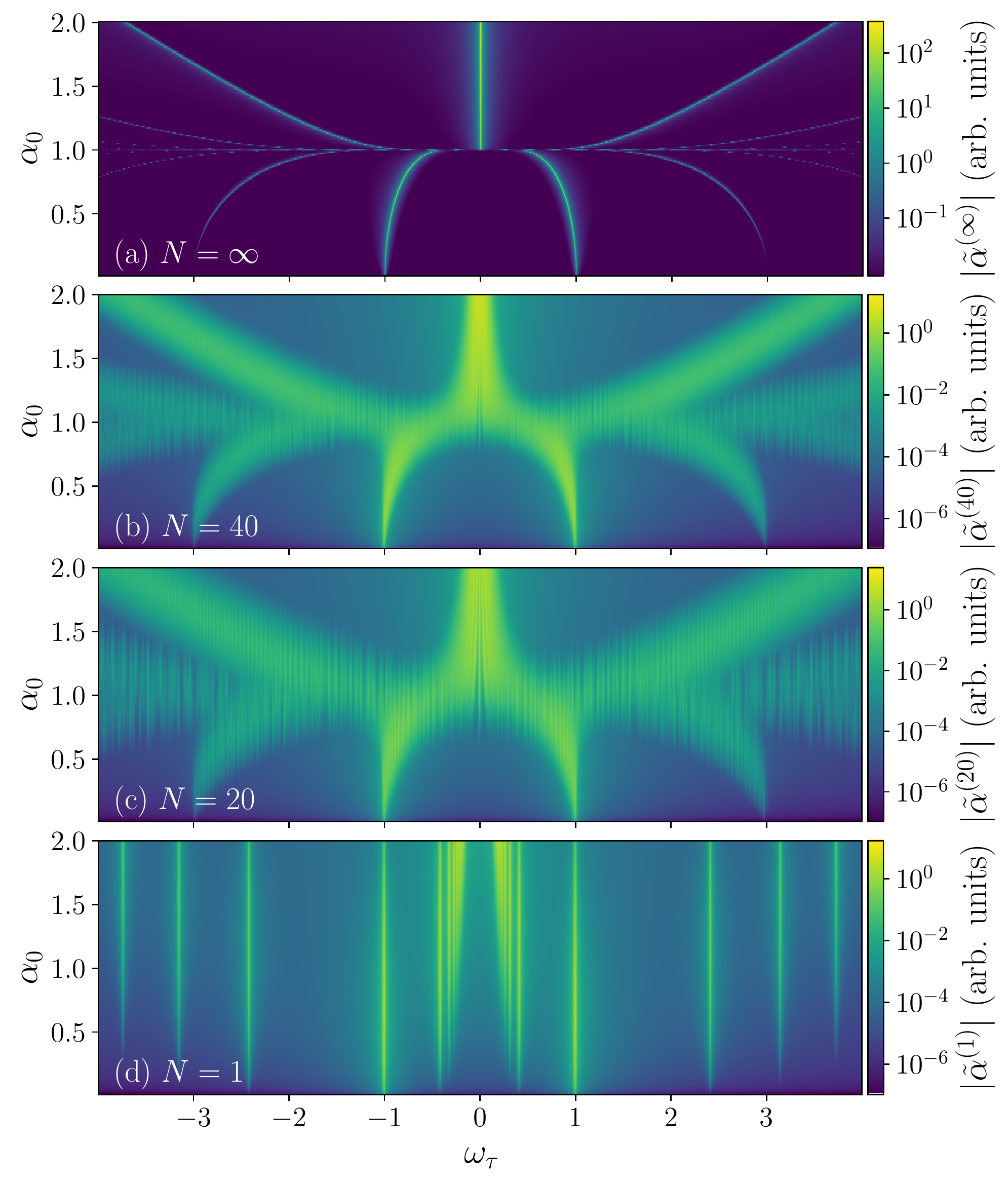}
 \caption{Spectrum of $\highindex{\alpha}{N}$ for different initial values $\alpha_0$. (a) Semiclassical limit determined via \cref{eq:eom_semiclassical} together with \cref{eq:fourier}. (b)-(d) Quantum optical treatment determined via \cref{eq:erw_x_N} together with \cref{eq:fourier} for the cases (b) $N=40$, (c) $N = 20$ and (d) $N=1$. The coupling strength is chosen, such that $\hbar G \sqrt{N} = const.$, independent of $N$.}
 \label{fig2}
\end{figure}

In \cref{fig2} we show the spectrum of the light field, i.e., the Fourier transform of $\alpha(\tau)$, depending on the initial coherent amplitude $\alpha_0$ for different numbers $N$ of QEs.
The spectrum $\tilde{f}(\omega_\tau)$ of a function $f(\tau)$ is defined by
\begin{align}
 \tilde{f}(\omega_\tau) = \Re \left(\int \dd{\tau} f(\tau) e^{-i\omega_\tau \tau - \epsilon \abs{\tau}}\right) \label{eq:fourier}
\end{align}
where $\omega_\tau$ is the frequency corresponding to the dimensionless time $\tau$ defined in \cref{eq:trafo_tau} and $\epsilon = 0.01\, \te{meV} / \hbar G \sqrt{N}$ has been introduced as a damping parameter to broaden the spectral distribution for a better visibility.

Let us start by briefly reviewing the result obtained in the semiclassical limit ($N = \infty$) for a $\delta$-shaped ensemble of QEs (see Ref.~\cite{jurgens2020sem}).
As shown in \cref{fig2} (a), we find exciton-polariton dynamics for initial values $\alpha_0 < 1$, which results in two spectral peaks at frequencies $\omega_\tau = \pm \omega_{\alpha_0}$ in the spectrum of the light field amplitude. For small initial values $\alpha_0 \ll 1$ the polariton frequency is given by $\omega_{\alpha_0}=1$, with increasing $\alpha_0$ it decreases. In this regime the occupation of the QEs never reaches one. 
For initial values $\alpha_0 > 1$ three peaks at $\omega_\tau = 0, \pm \omega_{\alpha_0}$ are found. This reflects Rabi oscillations with the Rabi frequency approaching  $\omega_{\alpha_0} = \sqrt{2(2\alpha_0^2 - 1)}$ for $\alpha_0 \gg 1$. The peak at the frequency zero reflects the fact that the field amplitude never vanishes in this regime.
At $\alpha_0 = 1$ a sharp transition between these regimes is observed.
We will call this initial value the transition point.
This transition can be explained by noting that \cref{eq:eom_semiclassical} can be transformed into the equation of motion for a classical particle moving in a double well potential which depends on the initial conditions of the dynamics \cite{jurgens2020sem}. Similar transitions to such three peak structures have also been found in the case of damped and driven systems \cite{munoz2019sym}.

Spectra obtained from the TC model for $N = 40, 20, 1$ are plotted in \cref{fig2} (b)-(d) again as functions of the initial value $\alpha_0$. For comparability we have set $\hbar G \sqrt{N} = const.$ in all calculations.
We apply this scaling of the coupling constant, such that the polariton splitting is independent of the number of QEs and can be easily compared to the semiclassical model.

\begin{figure}
 \centering
 \includegraphics[width=1\columnwidth]{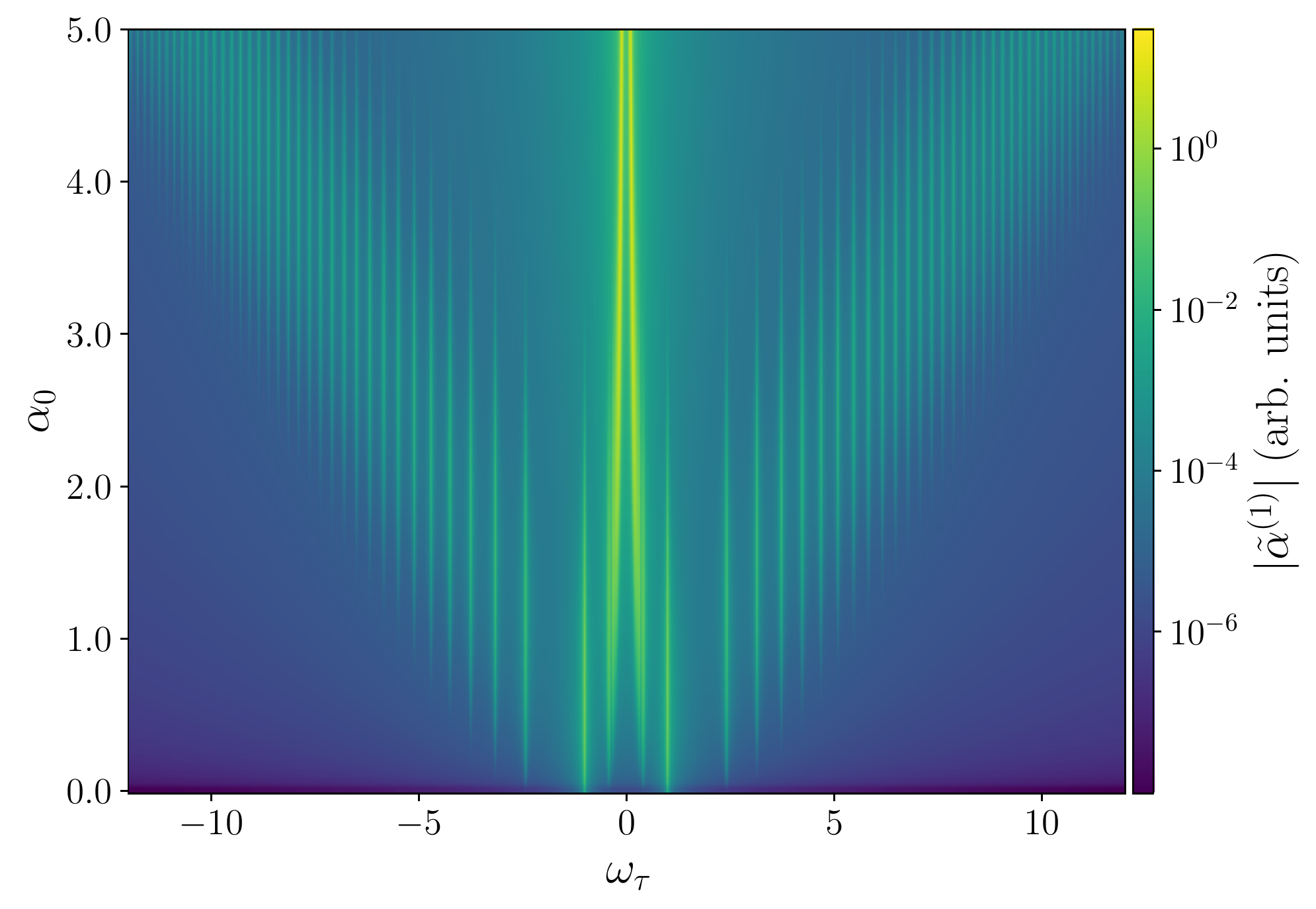}
 \caption{Spectrum of $\highindex{\alpha}{1}$ (i.e., $N=1$) for initial values between $\alpha_0 = 0$ and $\alpha_0 = 5$.}
 \label{fig3}
\end{figure}

For $N=1$ [\cref{fig2} (d)] we see the spectrum of the JC model.
Here the purely discrete nature of the JC ladder can be observed leading to energetically well separated transitions between the ladder rungs for the relatively low number of photons (note that $\alpha_0=2$ corresponds to an initial mean photon number of four).
At very small values $\alpha_0 \ll 1$ only the lowest rung of the JC ladder contributes to the dynamics leading to the same two polariton frequencies as in the semiclassical case, as discussed before.
For increasing $\alpha_0$, however, additional spectral lines appear.
In contrast to the semiclassical model shown in panel (a), the sharp transition at $\alpha_0 = 1$ cannot be observed and no continuous shift of the transition lines as a function of $\alpha_0$ is seen.
Also for initial values $\alpha_0 > 1$, where the semiclassical model reveals Rabi oscillations, the JC spectrum in this parameter range (\cref{fig2} (d)) looks very different from the semiclassical spectrum in (a).
If $\alpha_0$ is increased further, the JC ladder splittings of neighboring rungs approach each other and neighboring spectral lines get closer leading to essentially three lines reflecting the physics of the Mollow triplet \cite{mollow1969pow,valle2010mol} and thus approaching the same structure as in the semiclassical case.
This is shown in \cref{fig3}, where the spectrum obtained from the JC model is plotted for initial values up to $\alpha_0 = 5$ (corresponding to 25 photons on average).

In Figs.~\ref{fig2} (b) and (c) the spectra are shown for QE numbers $N=40$ and $N=20$, respectively.
For small initial values $\alpha_0 \ll 1$ the exciton-polariton splitting with peaks at $\omega_\tau= \pm 1$ can be seen independent of the number of QEs.
For higher initial values the number of eigenstates contributing to the spectra at a given value $\alpha_0$ strongly increases with increasing $N$, resulting in a decrease of the energy separation of neighboring spectral peaks.
For $N=20$  [\cref{fig2} (c)] the general structure of the semiclassical spectrum can already be recognized but the discrete structure of the quantized system is still well resolved.
At $N=40$  [\cref{fig2} (b)] the broadened $\delta$-peaks overlap and the general structure of the semiclassical case, shown in panel (a), is now clearly visible.
Compared to the semiclassical model (a), the transition at $\alpha_0 = 1$ is not as sharp and the discrete structure of the individual rungs can still be seen around $\alpha_0 = 1$.
In the semiclassical model the transition point $\alpha_0 = 1$ corresponds to an infinitely slow evolution of the system into an unstable fixed point.
When this fixed point is reached all QEs are excited and there are no more photons left inside the cavity \cite{jurgens2020sem}.
Since quantum effects, in particular spontaneous emission, are not included in the semiclassical model, the system stays in its unstable fixed point.
In the region around $\alpha_0 = 1$ quantum effects play an important role, leading to the broadened transition seen in \cref{fig2}.
These findings reflect the strong influence of the QE-photon correlations at $\alpha_0 \approx 1$ similar to Ref.~\cite{zens2019cri}.

Although the sharp transition at $\alpha_0=1$ is not present in the quantum model, many features of the semiclassical limit can already be seen for relatively small $N\approx 40$.
In the following we will compare the dynamics obtained from the semiclassical and the quantum optical calculations to get a better understanding of the differences between these two approaches.

\subsection{Dynamics}
\begin{figure}
 \centering
 \includegraphics[width=1\columnwidth]{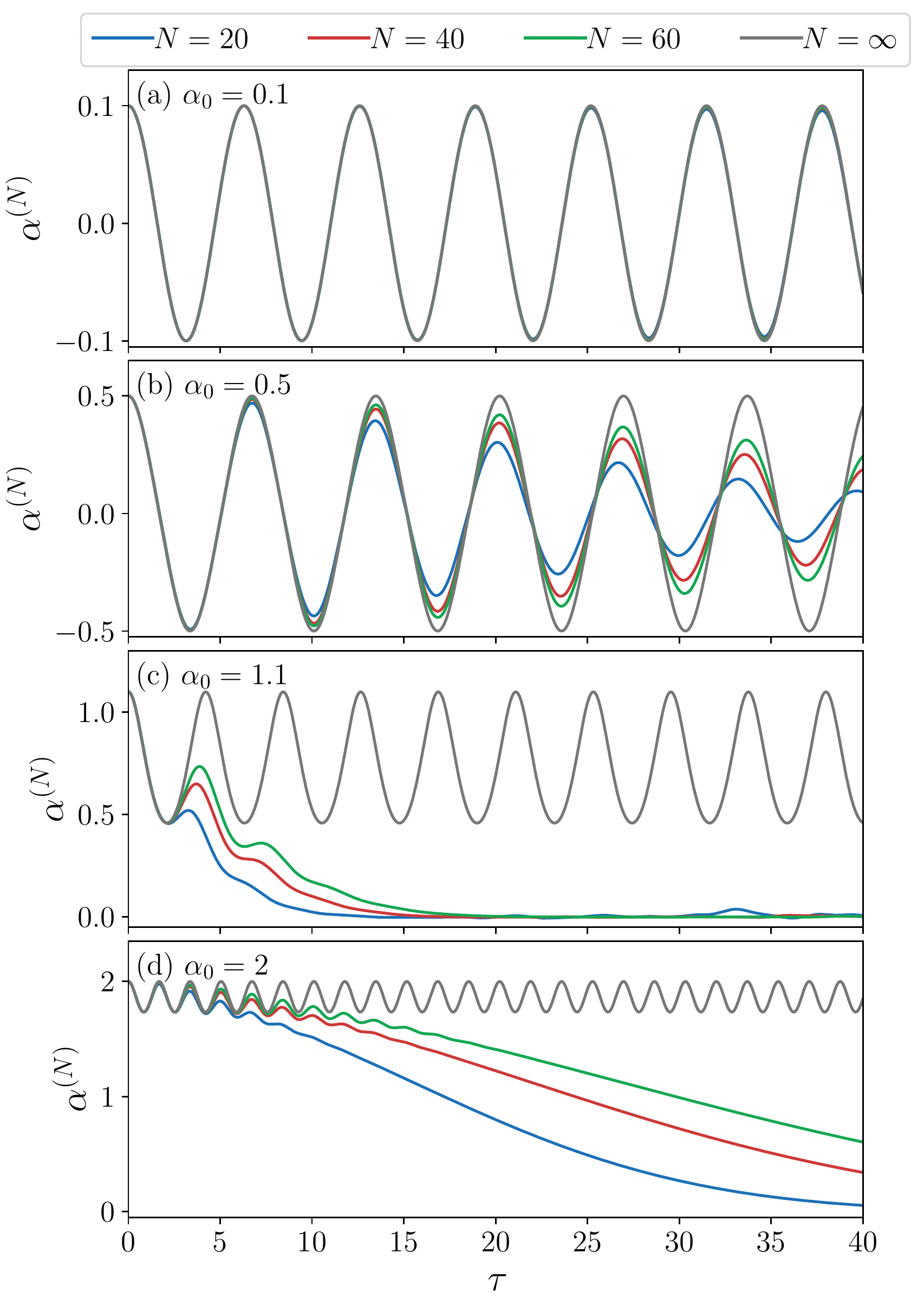}
 \caption{Dynamics of $\highindex{\alpha}{N}$ for different initial conditions (a) $\alpha_0 = 0.1$, (b) $\alpha_0 = 0.5$, (c) $\alpha_0 = 1.1$ and (d) $\alpha_0 = 2$ and different numbers of QEs [$N=20$ (blue line), $N=40$ (red line), $N=60$ (green line), $N=\infty$ (grey line)]. The oscillation periods in the case $N=\infty$ are found to be $\tau_{0.1}\approx 6.28$, $\tau_{0.5}\approx 6.7$, $\tau_{1.1}\approx 4.1$, and $\tau_{2.0} \approx 1.68$.}
 \label{fig4}
\end{figure}

The dynamics of $\highindex{\alpha}{N}$ are shown in \cref{fig4} for different initial values $\alpha_0$ and different $N$ as a function of the dimension-less time $\tau$ [\cref{eq:trafo_tau}].
Note that while in general $\highindex{\alpha}{N}$ and $\highindex{\alpha}{\infty}$ are complex variables, in the case of resonance ($\omega_x = \omega_0$) between the light field and the QE transitions as well as for real initial conditions, as considered here, they are real quantities.
This also explains the symmetry with respect to $\omega=0$ in all the spectra in \cref{fig2,fig3}.

In the semiclassical model (grey line) the coherent field amplitude exhibits undamped oscillations for all values of $\alpha_0 \ne 1$ with an oscillation period given by $\tau_{\alpha_0}=2\pi/\omega_{\alpha_0}$. For the lowest and highest values of $\alpha_0$ the oscillation periods essentially agree with the polaritonic limit $\tau_{0.1}\approx 2\pi$ and the Rabi frequency limit $\tau_{2.0} \approx 2\pi / \sqrt{2(2\alpha_0^2 - 1)} \approx 1.68$. For the intermediate values we find $\tau_{0.5}\approx 6.7$ and $\tau_{1.1}\approx 4.1$. 

For small values [$\alpha_0=0.1$ in \cref{fig4} (a)] the lines of the quantum optical (colored lines) and semiclassical solutions cannot be distinguished, thereby showing an almost perfect agreement. 
For larger initial values [$\alpha_0=0.5$ in \cref{fig4} (b)] $\highindex{\alpha}{N}$ also agrees with the semiclassical model for small $\tau$.
However, for times $\tau \gtrsim 10$ the deviations between the quantum calculations and the semiclassical model become more pronounced; furthermore they are stronger for smaller numbers of QEs, as can be seen by comparing the case of $N=20$ (blue line) with the cases $N=40$ (red line) and $N=60$ (green line).
Interestingly, we find that the frequencies of the oscillations fit very well, even if the amplitudes increasingly differ as a function of time, which explains the qualitatively good agreement of the spectra [\cref{fig2} (a)-(c)] in terms of the positions of the peaks.
At initial values close to $\alpha_0=1$, where in the semiclassical model the transition occurs [\cref{fig4} (c), $\alpha_0=1.1$], the dynamics differ strongly already after the first minimum of the coherent amplitude, because the correlations have a strong impact in this parameter regime, as has been shown in Ref.~\cite{zens2019cri} in the damped and driven system.
For still larger initial values [$\alpha_0=2$ in \cref{fig4} (d)], we see that the dynamics agree for the first few oscillations and then increasingly deviate.
Again, the deviations start earlier and are stronger for smaller QE numbers $N$. 
The strong decay seen in \cref{fig4} (d) reflects the collapse part of the collapse and revival phenomenon in the JC and TC model \cite{ramon1998col,agarwal2012tav,meunier2006ent,jarvis2009dyn}.

Thus we can conclude from \cref{fig4} that for larger $N$ (green curves) the coherent amplitude agrees with the semiclassical model on longer time scales than for smaller $N$ (red and blue curves). 
Apart from the dynamical transition around $\alpha_0 = 1$, the frequency in general matches very well, resulting in the good agreement of the position of the resonances within the spectra, even if the amplitudes differ strongly for larger times. 
The large broadening of the peak structure seen in \cref{fig2} (b) and (c) is caused by the collapse of the coherent amplitude; indeed we find that the spectra are less broadened for higher numbers of QEs, where the collapse occurs at later times. 
A larger collapse time therefore leads to sharper peaks in the spectrum and a better matching of the dynamics between the semiclassical and quantum optical approach.

\begin{figure}[t]
 \centering
 \includegraphics[width=0.9\columnwidth]{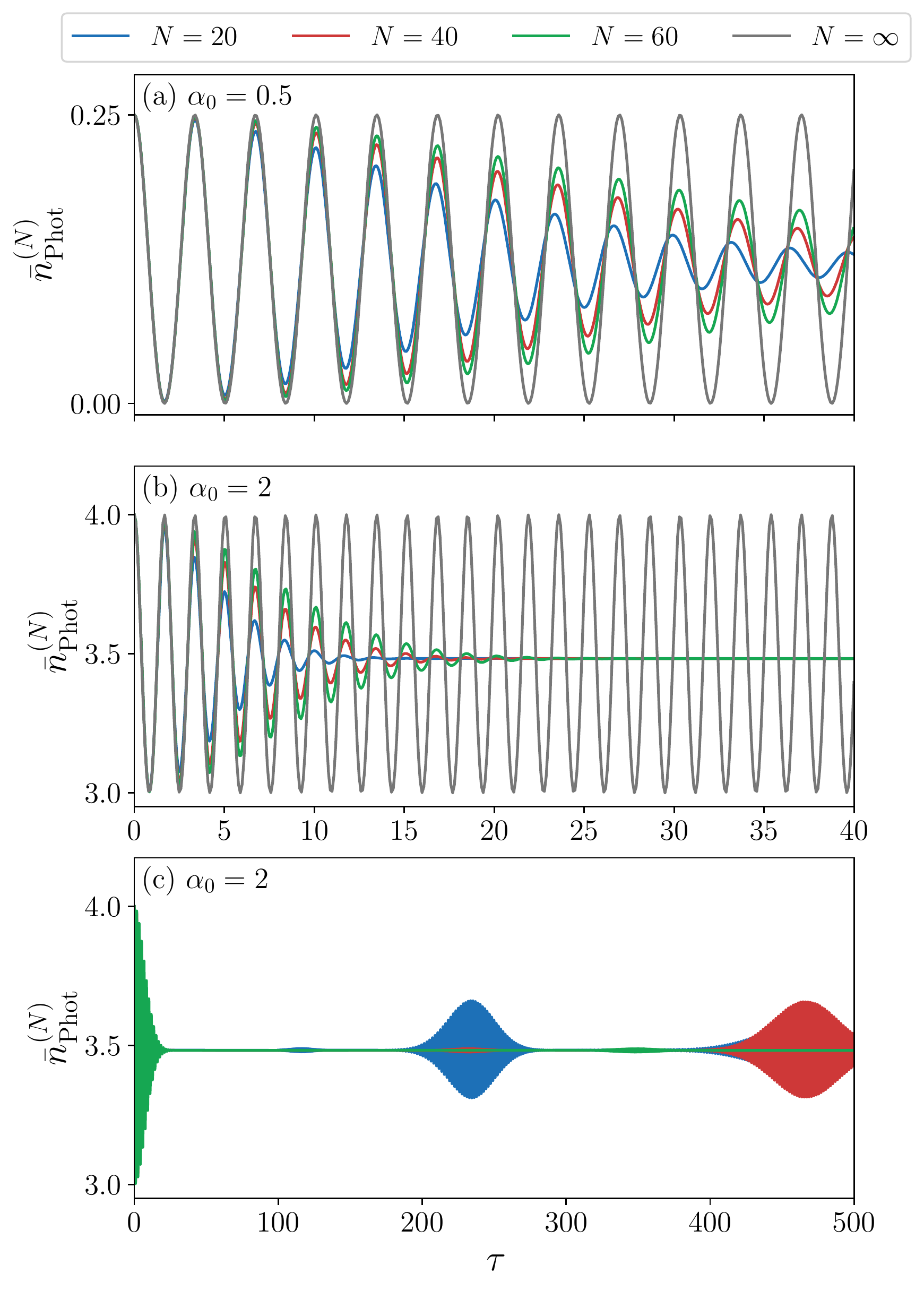}
 \caption{Dynamics of the mean photon number per QE $\bar{n}_{\te{Phot}}^{(N)} = \erw{\hat{\alpha}^\dagger \hat{\alpha}}$ for (a) $\alpha_0 = 0.5$, (b) and (c) $\alpha_0 = 2$ for different numbers of QEs $N$. Panel (c) shows the same results as in (b) for $N=20$, $40$, $60$ on a longer time scale.}
 \label{fig5}
\end{figure}

While in the semiclassical model, due to the coherence of the light field, the field intensity is strictly related to the field amplitude, this does not hold in the quantum optical model anymore.
In fact, the collapse and revival phenomenon in JC and TC models is associated with the loss and build-up of coherence of the light field, respectively.
To get a deeper understanding of these phenomena in our system we investigate the dynamics of the mean number of photons per QE
\begin{align}
\overline{n}_\te{Phot}^{(N)}(t) = \bra{\highindex{\Psi}{N}(t)}\hat{\alpha}^\dagger \hat{\alpha}\ket{\highindex{\Psi}{N}(t)}.
\end{align}
$\overline{n}_\te{Phot}^{(N)} (t)$ is calculated analog to \cref{eq:erw_x_N}.
The total number of photons in the system is then given by $N \cdot \overline{n}_\te{Phot}^{(N)} $.
Initially the light field is in a coherent state with $\overline{n}_\te{Phot}^{(N)}(0) = \abs{\alpha_0}^2$.
In the semiclassical model the system remains in a coherent state, such that at any time the relation $\overline{n}_\te{Phot}^{(\infty)}(t) = \abs{\highindex{\alpha}{\infty}(t)}^2$ holds.

The dynamics of the mean photon number per QE for different $\alpha_0$ and $N$ is shown in \cref{fig5}, where the line colors have the same meaning as in \cref{fig4}.
For small initial values ($\alpha_0 = 0.5$) the temporal evolution of the mean photon number is shown in \cref{fig5} (a).
In this case there are initially on average $0.25$ photons per QE in the system, i.e., there are less photons than QEs. 
Every QE can thus absorb and re-emit up to $0.25$ photon on average.
In the semiclassical model this happens periodically and $\overline{n}_\te{Phot}^{(\infty)}$ continues to oscillate between $0.25$ and zero.
In the quantum optical model the oscillations decay and, as the coherent amplitude in \cref{fig4} (b) relaxes to zero, the mean photon number relaxes to approximately half the initial value.
The dynamics of the mean photon number above the transition point (here $\alpha_0 = 2$) are shown in \cref{fig5} (b) and the respective long time behavior is shown in \cref{fig5} (c).
Now, initially the mean photon number per QE is four.
Each QE absorbs and re-emits up to one photon leading to the oscillations of $\overline{n}_\te{Phot}$ between $4$ and $3$.
This corresponds to modulations between $4N$ and $3N$ photons in total.
Again, the oscillations persist in the semiclassical model while in the quantum optical case at longer times a quasi-stationary value of $\overline{n}_\te{Phot}^{(N)} = 3.5$ is reached, indicating that every QE has absorbed on average half a photon.
In the long time behavior [see \cref{fig5} (c)] besides the just described collapse also the corresponding revival can be seen \cite{agarwal2012tav,jarvis2009dyn} (note that for $N=60$ the revival is beyond the plotted times).
With the given initial conditions the revival time increases with the number of QEs, as seen in \cref{fig5} (c).
We will come back to the phenomenon of collapse and revival in Sec.~\ref{sec.collapse}. 

\section{Wigner functions}\label{sec.wigner}
The different time evolutions of the squared coherent field amplitude and of the mean photon number are clear indications that the quantum state of the light field strongly deviates from a coherent state.
In the following we will analyze these deviations in more detail.
The complete quantum state of a single mode light field can be best visualized in a phase space representation by calculating the corresponding Husimi $Q$ function or the Wigner function.
We focus here on the Wigner function due to the easy identification of non-classical features in terms of negativities of this function.
The Wigner function of a coherent state is an isotropic two-dimensional Gaussian.
Deviations from the coherent state are thus directly reflected in deformations of the Wigner function from this shape.

The Wigner function $\wigner{W}{N}{\lambda, t}$ is a function of the complex phase space variable $\lambda = U + i\Pi$ and defined by \cite{haroche2006exp}
\begin{align}
 \wigner{W}{N}{\lambda, t} = \frac{2}{\pi} \Tr[ D(-\lambda) \hat{\rho}^{(N)}(t) D(\lambda) \mathcal{P} ]
\end{align}
with the density operator $\hat{\rho}^{(N)}(t) = \kb{\Psi^{(N)}(t)}{\Psi^{(N)}(t)}$, the displacement operator $D(\lambda) = \exp(\lambda \hat{a}^\dagger - \lambda^* \hat{a})$, the parity operator $\mathcal{P} = \exp(i\pi \hat{a}^\dagger \hat{a})$, and the trace $\Tr[\dots]$.
The superscript $(N)$ denotes the number of QEs in the system.
Using the eigenstates $\ket{j, n; l}$ and the state $\ket{\Psi^{(N)}(t)}$ defined in \cref{eq:time_evolution_state}, the Wigner function can be written as 
\begin{align}
  \wigner{W}{N}{\lambda, t} = \sum_{\substack{n, n' \\l, l'}} \rho^{(N,n, n')}_{l, l'}(t) \, w^{(N,n, n')}_{l, l'}(\lambda) \nonumber
\end{align}
with 
\begin{align}
w^{(N,n, n')}_{l, l'}(\lambda) &= \frac{2}{\pi} \Tr[D(-\lambda) \kb{j, n; l}{j, n'; l'} D(\lambda) \mathcal{P}] , \nonumber \\
\rho^{(N, n, n')}_{l, l'}(t) &= \exp(-\abs{\xi}^2) \frac{\xi^n}{\sqrt{n!}} \frac{\xi^{*n'}}{\sqrt{n'!}} \ccoeff{c}{0, l}{N,n} \coeff{c}{0, l'}{N, n'} \nonumber \\
&\times \exp[-\frac{i}{\hbar} \left( E_l^{(N,n)} - E_{l'}^{(N,n')} \right) t ]. \nonumber
\end{align}
As is shown in  \cref{app:wigner}, by using the properties of the Fock states, a closed expression for the Wigner function can be obtained.

To relate the Wigner function to our results, we express it in the transformed frame defined by \cref{eq:transformation}.
By introducing the coordinates in the transformed frame $U_\alpha = U/\sqrt{N}$ and $\Pi_\alpha = \Pi/\sqrt{N}$ and the transformed Wigner function 
\begin{align}
&\wigner{\widetilde{W}}{N}{U_\alpha, \Pi_\alpha, \tau} \nonumber\\
&\quad = N \, \wigner{W}{N}{\sqrt{N} (U_\alpha + i\Pi_\alpha) e^{-i\omega_0 \frac{\tau}{g\sqrt{N}}}, \frac{\tau}{g\sqrt{N}}},
\end{align}
the field amplitude is obtained by
\begin{align}
\highindex{\alpha}{N} &= \erw{\hat{\alpha}}^{(N)}  \\
&=\int \dd{U_\alpha} \dd{\Pi_\alpha} \left( U_\alpha + i\Pi_\alpha  \right) \wigner{\widetilde{W}}{N}{U_\alpha, \Pi_\alpha, \tau},  \label{eq:erw_wigner} \nonumber
\end{align}
with $\hat{\alpha} = \hat{\tilde{a}} / \sqrt{N}$.
The derivation can be found in \cref{app:wigner_transformation}.
The quadrature operators in the transformed frame are given by
\begin{align}
 \hat{u}_\alpha = \frac{1}{2} \left( \hat{\alpha} + \hat{\alpha}^\dagger \right), \hspace{1cm}
 \hat{\pi}_\alpha = \frac{1}{2i} \left( \hat{\alpha} - \hat{\alpha}^\dagger \right). \nonumber
\end{align}
As already mentioned above, in all the cases studied here $\highindex{\alpha}{N}$ is a real quantity leading to 
\begin{subequations}
\begin{align}
\erw{\hat{u}_\alpha}^{(N)} &= \highindex{\alpha}{N}, \\
\erw{\hat{\pi}_\alpha}^{(N)} &= 0. 
\end{align}
\end{subequations}
This reflects the fact that the transformed Wigner function is here always symmetric in $\Pi_\alpha$.

The variances of the Wigner function are given by
    \begin{align}
    &\left(\Delta u_{\alpha}^{(N)}\right)^2 = \erw{\hat{u}_\alpha^2}^{(N)} - \left(\erw{\hat{u}_\alpha}^{(N)}\right)^2,  \nonumber \\
    &\left(\Delta \pi_{\alpha}^{(N)}\right)^2 = \erw{\hat{\pi}_\alpha^2}^{(N)} - \left(\erw{\hat{\pi}_\alpha}^{(N)}\right)^2, \nonumber
    \end{align}
where $\Delta u_{\alpha}^{(N)}$ and $\Delta \pi_{\alpha}^{(N)}$ are the widths in the $U_\alpha$ and $\Pi_\alpha$ direction, respectively.
At $\tau = 0$ the Wigner function is given by a Gaussian distribution shifted in $U_\alpha$ direction by the amount $\alpha_0$ and with widths $\Delta u_{\alpha, \te{coh.}}^{(N)} = \Delta \pi_{\alpha, \te{coh.}}^{(N)} = (4N)^{-1/2}$ in the scaled frame.
Deviations from this value indicate nonclassical states, because they are directly related to the incoherent part of the photon occupation $\erw{\alpha^\dagger \alpha}^{(N)}_{\mathrm{incoh.}}$ according to
\begin{subequations}\label{eq:fluctuations}
    \begin{align}
    \left(\Delta_u^{(N)}\right)^2 &= \left( \Delta u_\alpha^{(N)} \right)^2 - \left( \Delta u_{\alpha, \te{coh}}^{(N)} \right)^2, \\
    \left(\Delta_\pi^{(N)}\right)^2 &= \left( \Delta \pi_\alpha^{(N)} \right)^2 - \left( \Delta \pi_{\alpha, \te{coh}}^{(N)} \right)^2, \\
    \erw{\hat{\alpha}^\dagger \hat{\alpha}}_{\mathrm{incoh.}}^{(N)} &= \erw{\hat{\alpha}^\dagger \hat{\alpha}}^{(N)} - \erw{\hat{\alpha}^\dagger}^{(N)} \erw{\hat{\alpha}}^{(N)} \nonumber \\
    &= \left( \Delta_u^{(N)} \right)^2 + \left( \Delta_\pi^{(N)} \right)^2\, .
    \end{align}
\end{subequations}
\noindent
We will refer to $\Delta_u^{(N)}$ and $\Delta_\pi^{(N)}$ as excess fluctuations.

\begin{figure}
 \centering
 \includegraphics[width=1\columnwidth]{{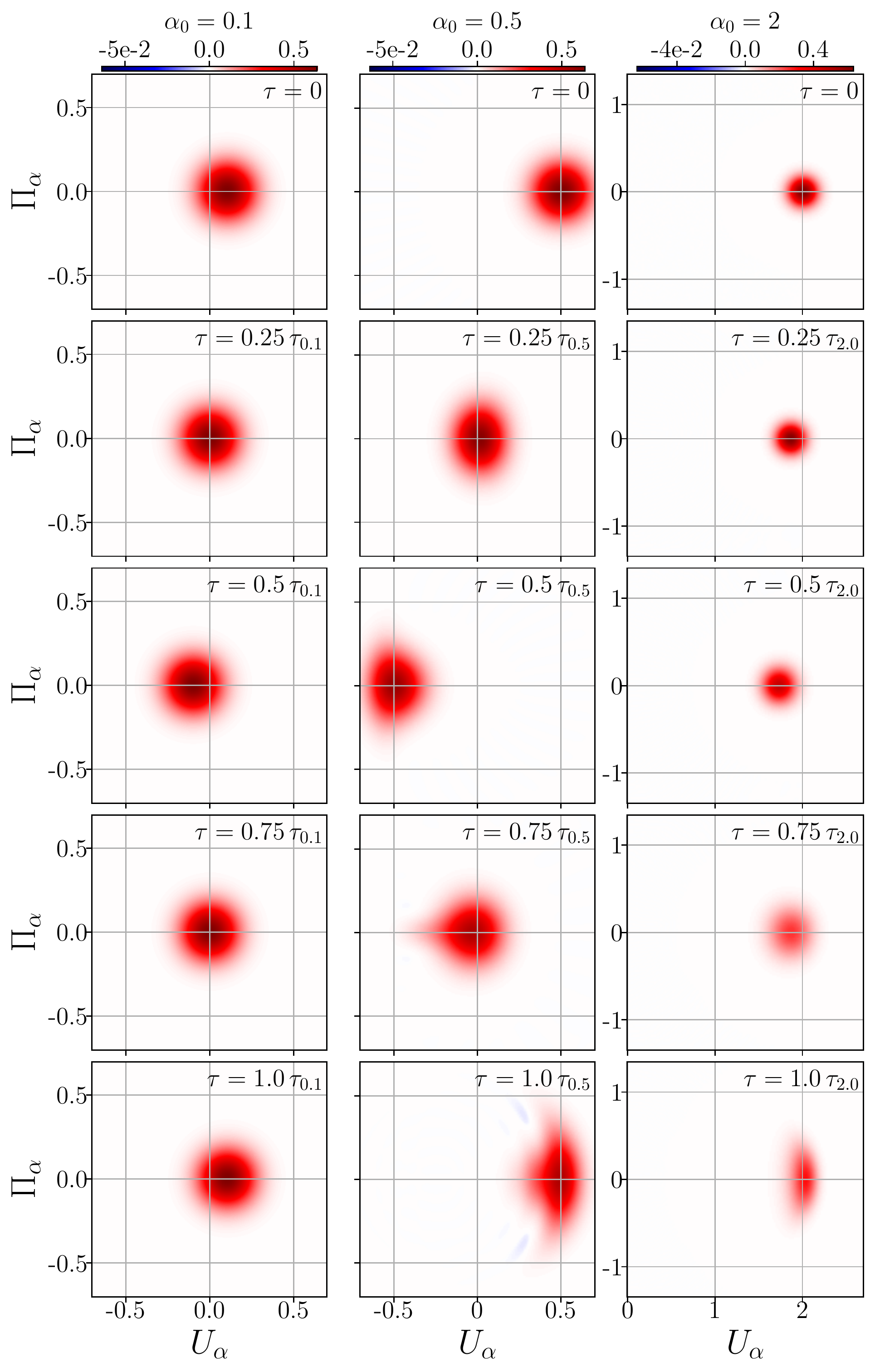}}
 \caption{Wigner functions $\wigner{\widetilde{W}}{N}{U_\alpha, \Pi_\alpha, t}/N$ for initial values $\alpha_0 = 0.1$ (left column), $\alpha_0 = 0.5$ (central column) and $\alpha_0 = 2$ (right column) for $N=20$ QEs at different times during the first oscillation period $0 \leq \tau \leq \tau_{\alpha_0}$ (increasing from top to bottom). The oscillation periods are $\tau_{0.1}\approx 6.28$, $\tau_{0.5}\approx 6.7$, and $\tau_{2.0} \approx 1.68$ (see \cref{fig4}). Note that the axes of the right column are scaled differently.}
 \label{fig6}
\end{figure}

We now discuss the Wigner function $\wigner{\widetilde{W}}{N}{U_\alpha, \Pi_\alpha, \tau}$ at the extrema and turning points of $\highindex{\alpha}{N}(\tau)$ during the first period.
The Wigner function at these times is shown for $N=20$ QEs in \cref{fig6} for the three initial values $\alpha_0 = 0.1$ (left column), $\alpha_0 = 0.5$ (central column), and $\alpha_0 = 2.0$ (right column).
The times, increasing from top to bottom, are given in units of the intrinsic oscillation period $\tau_{\alpha_0}$, which for the present cases have been found to be $\tau_{0.1}\approx 6.28$, $\tau_{0.5}\approx 6.7$, and $\tau_{2.0} \approx 1.68$ (see \cref{fig4}).
Note that the color scale is chosen such that saturation is reached for smaller values in the negative than positive direction to easier identify negativities in the Wigner function, which are clear signatures of quantum behavior.

For initial values $\alpha_0 \leq 1$ (first two columns) the Wigner function oscillates between $+\alpha_0$ and $-\alpha_0$ on the $\Pi_\alpha = 0$ line in $U_\alpha$-direction.
Note that the untransformed Wigner function $\wigner{W}{N}{\lambda, t}$ rotates with the frequency $\omega_0$ around the origin, while the transformed function $\wigner{\widetilde{W}}{N}{U_\alpha, \Pi_\alpha, \tau}$ is defined in this rotating frame and therefore oscillates along the $U_\alpha$ axis.

For $\alpha_0 = 0.1$ (left column) the Wigner function keeps its Gaussian shape during the first period and it essentially returns to the initial state after a complete oscillation.
This is in agreement with the previous finding that the system behaves similar to the semiclassical one.
For $\alpha_0 = 0.5$ (central column) the shape of the Wigner function changes and after a complete oscillation ($\tau = \tau_{0.5}$, bottom panel) we already see clear deviations from the initial state associated with very small negative contributions.
In addition to the negative parts, the distribution broadens mainly in the $\Pi_\alpha$ direction, while the width in the $U_\alpha$ direction stays approximately the same as in the initial state.
Using \cref{eq:erw_wigner}, one can see that the broadening in the $\Pi_\alpha$ direction has no impact on the real part of the expectation value.
Therefore, the mean value $\erw{u_\alpha}^{(N)}(\tau)$ does not deviate strongly from the coherent case at the same time, which results in the good agreement with the semiclassical model, as seen in \cref{fig4}.
Despite this similarity of the mean value with the coherent amplitude, we see clear changes in the quantum state.

For $\alpha_0 = 2$ (right column), i.e., above the transition point, the dynamics during the first period are restricted to positive $U_\alpha$, reflecting the fact that now $\highindex{\alpha}{N}$ remains positive (see \cref{fig4}).
Again, after the first oscillation the Wigner function is stretched in the $\Pi_\alpha$ direction compared to the initial state.
Similar to the case with $\alpha_0 = 0.5$, the distribution in the $U_\alpha$ direction does not differ significantly, leading to the good agreement of the dynamics of the mean value $\erw{\hat{u}_\alpha}^{(N)}(\tau)$ during the first oscillation period, as seen in \cref{fig4}.

\begin{figure}
 \centering
 \includegraphics[width=1\columnwidth]{{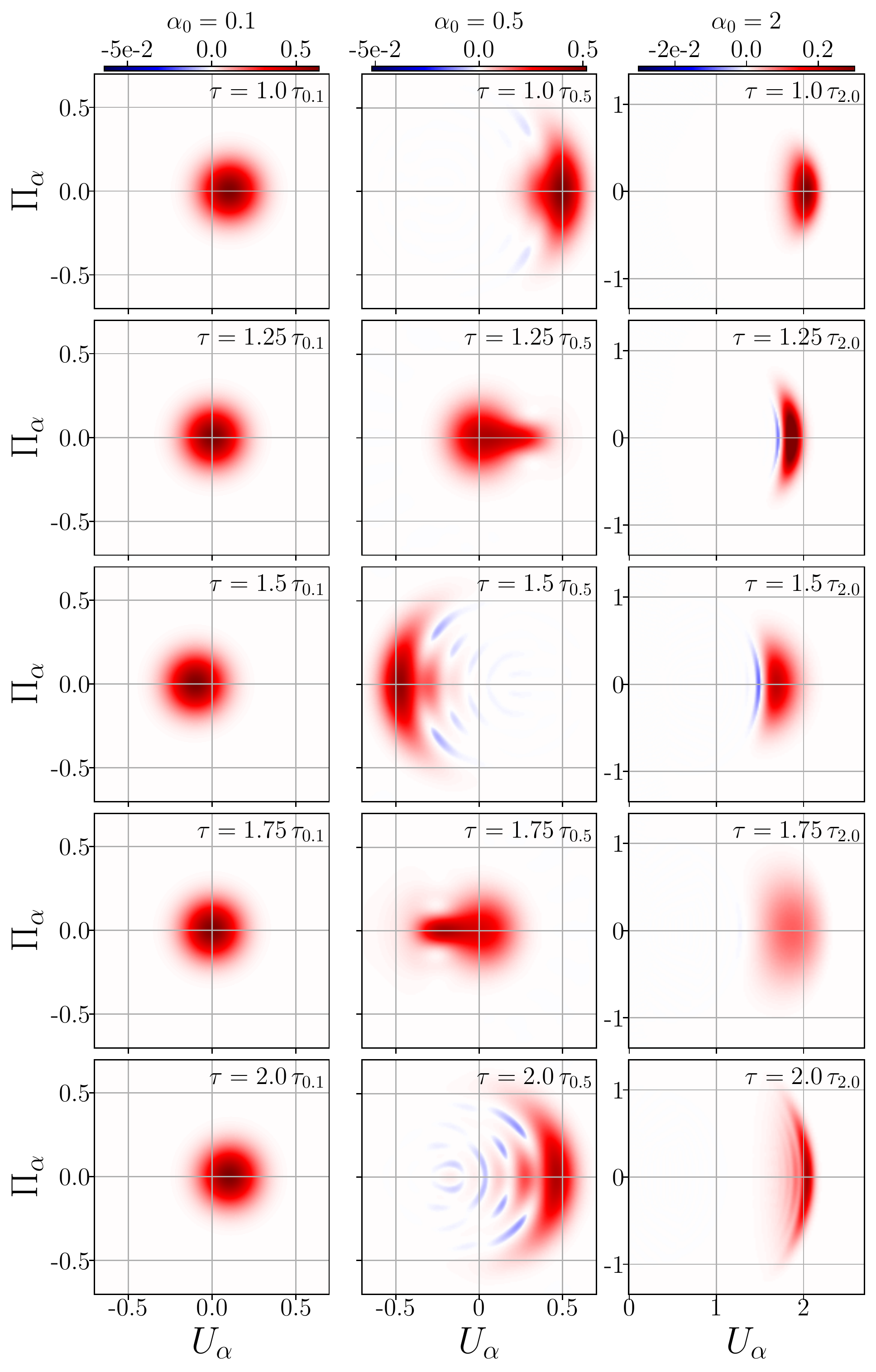}}
 \caption{Wigner functions $\wigner{\widetilde{W}}{N}{U_\alpha, \Pi_\alpha, t} / N$ for initial values $\alpha_0 = 0.1$ (left column), $\alpha_0 = 0.5$ (central column) and $\alpha_0 = 2$ (right column) for $N=20$ QEs at different times during the second oscillation period $\tau_{\alpha_0} \leq \tau \leq 2\tau_{\alpha_0}$ (increasing from top to bottom). The oscillation periods are $\tau_{0.1}\approx 6.28$, $\tau_{0.5}\approx 6.7$, and $\tau_{2.0} \approx 1.68$ (see \cref{fig4}). Note that the axes of the right column are scaled differently.}
 \label{fig7}
\end{figure}

The Wigner functions during the second oscillation period of \cref{fig4} are shown in \cref{fig7} for the same initial values.
Note that we changed the color bar scaling for $\alpha_0 = 2$ (right column) for better visibility.
For small initial values (left column) the same behavior as before continues.
This underlines the interpretation, that the semiclassical model is a good approximation for small initial values of the light field.
In this case the quantum state of the light remains to a large degree in a coherent state.
For larger amplitudes, but still below the transition ($\alpha_0 = 0.5$, central column), the deformations of the Wigner function further increase compared to the first period (central column, \cref{fig6}).
At the times $\tau=1.25\tau_{0.5}$ and $\tau=1.75\tau_{0.5}$ we now also observe a stretching along the $U_\alpha$ direction while for $\tau=1.5\tau_{0.5}$ and $\tau=2\tau_{0.5}$ the stretching occurs mainly in the $\Pi_\alpha$ direction.
At the latter times we also observe the development of a ringlike structure, similar to negativities appearing in the Wigner function of higher Fock states \cite{schleich2011qua}.
Here, however, the rings are not centered around the origin.
These times ($\tau = 1.5\tau_{0.5}$ and $\tau = 2\tau_{0.5}$) correspond to local extrema in the $\highindex{\alpha}{N}(t)$ dynamics, where the strongest deviations between quantum and semiclassical treatment have been observed in \cref{fig4}.
The same trend can be seen for $\alpha_0 = 2$ (right column), where the Wigner function exhibits a significant spread in the $\Pi_\alpha$ direction and negative parts at certain times.
Interestingly, here the negative parts are most prominent at times $\tau = 1.25\tau_{2.0}$ and $\tau=1.5\tau_{2.0}$, where $\highindex{\alpha}{N}$ is in good agreement with the semiclassical model (see \cref{fig4}).

\begin{figure}
 \centering
 \includegraphics[width=1\columnwidth]{{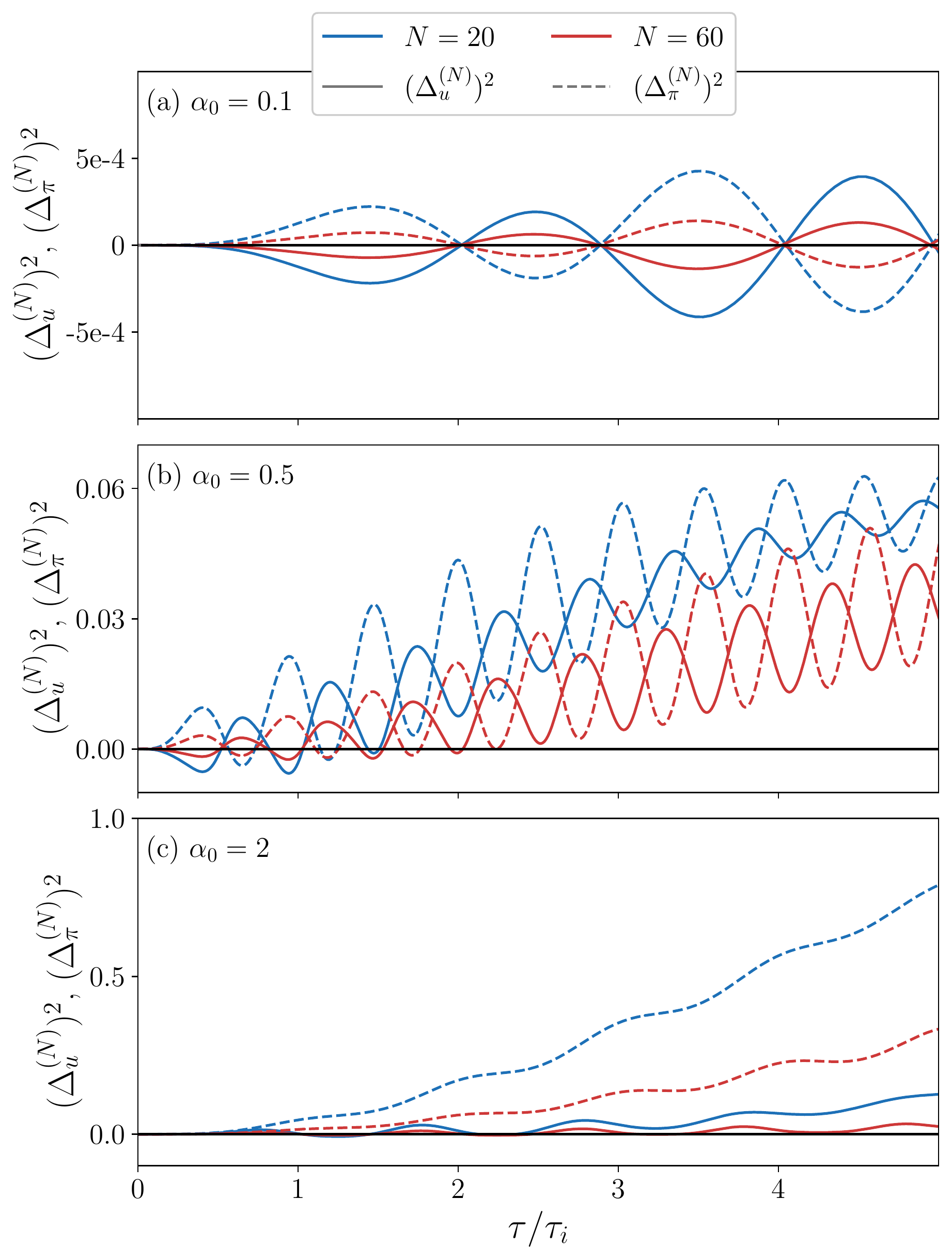}}
 \caption{Excess fluctuations $\Delta_u^{(N)}$ (solid) and $\Delta_\pi^{(N)}$ (dashed) of the Wigner function for $N=20$ (blue lines), $N=60$ (red lines) and $N=\infty$ (black lines).}
 \label{fig8}
\end{figure}

The color plots of the Wigner functions in \cref{fig6,fig7} provide a good overview of the qualitative changes of the quantum state of the light field compared to the coherent state in the semiclassical model.
However, it is often not so easy to extract quantitative information directly from these plots.
Therefore, in \cref{fig8} we investigate the temporal evolution of the excess fluctuations $\Delta_u^{(N)}$ (solid lines) in the $U_\alpha$ direction and $\Delta_\pi^{(N)}$ (dashed lines) in the $\Pi_\alpha$ direction, as defined in \cref{eq:fluctuations} (a) and (b) for $N=20$ (blue lines) and $N=60$ (red lines).

In the semiclassical limit the light field is in a coherent state, leading to $\Delta_u^{(\infty)} = \Delta_\pi^{(\infty)} = 0$ (black solid lines).
For finite numbers $N$ and for a small initial condition $\alpha_0 = 0.1$  [\cref{fig8} (a)] the fluctuations oscillate with a very small amplitude around the semiclassical value, indicating that already at this initial condition there are some deviations from the coherent state.
However, they are very small and therefore not visible in the color plots in \cref{fig6,fig7}.
For larger initial conditions as seen in \cref{fig8} (b) and (c), the excess fluctuations keep oscillating with a significant amplitude and additionally experience a steady increase.
While in the case of  $\alpha_0 = 0.5$ [\cref{fig8} (b)] the increase is similar for $\Delta_u^{(N)}$ and $\Delta_\pi^{(N)}$, for $\alpha_0 = 2$ [\cref{fig8} (c)] $\Delta_\pi^{(N)}$ grows much faster than $\Delta_u^{(N)}$, consistent with the finding of a large broadening in $\Pi_\alpha$ direction in \cref{fig7}.
We also see that the excess fluctuations during the first oscillation periods drop slightly below zero indicating squeezing \cite{retamal1997sqe,hassan1993per}.
While for $\alpha_0=0.1$ and $\alpha_0 = 0.5$ squeezing occurs both in the $U_\alpha$ and the $\Pi_\alpha$ direction, for $\alpha_0 = 2$ the increase of $\Delta_\pi^{(N)}$ is so strong that only $\Delta_u^{(N)}$ exhibits squeezing.
The formation of squeezed states on small time scales in the TC model has also been investigated for different initial states in Refs.~\cite{retamal1997sqe,seke1995squ,seke1997eff}.
Squeezing has also been observed in the resonance fluorescence of a single QD \cite{schulte2015qua}.
When comparing the fluctuations for different numbers of QEs, we find that the deviations from the value of the coherent state decrease with increasing $N$.
This is in line with the findings that the correlations vanish for large $N$ \cite{zens2019cri}, however it also shows that the values up to $N=60$ studied here are still far from this limiting case.

\section{Collapse and revival}\label{sec.collapse}
In the results discussed until now we have mainly focused on the initial part of the dynamics, in which the coherent amplitude decays and the oscillating mean photon number relaxes to a time-independent value.
As is well known for the JC and TC models, and as has already been seen in \cref{fig5}(c), this collapse of the oscillatory dynamics is accompanied by a revival of the coherent amplitude and of the oscillations in the mean photon number at a revival time $\tau_R$.
For a fixed and sufficiently large initial value $\alpha_0 = \xi / \sqrt{N}$ and with $\hbar G \sqrt{N}= \te{const.}$ the revival time $\tau_{R,\te{analytical}}=2\pi \alpha_0 N$ (compare Ref.~\cite{meunier2006ent}) increases linearly with increasing $N$.
Note that by choosing $\xi = \te{const.}$ and $g=\te{const.}$ the revival time would be independent of the number of QEs.

For the JC model it has been shown that in the limit of large mean photon numbers the field and the QE states factorize at half-integer and integer multiples of the revival time \cite{gea-banacloche1990col,gea-banacloche1991ato}.
While at integer multiples the field is again in a coherent state, for half-integer multiples it is in a superposition of two coherent states, i.e., in a Schr\"odinger cat state.
It has been shown that in the case of a single QD in a cavity this can be used for the preparation of a cat state \cite{cosacchi2021sch}.
These results have been generalized to the TC model \cite{meunier2006ent,jarvis2009dyn}.
There, however, a complete factorization occurs only for a specific class of initial states, which does not comprise the initial state in our case, consisting of all QEs in their respective ground state.
Nevertheless, as will be seen below, the field exhibits interesting quantum states at specific times.

\begin{figure}
 \centering
 \includegraphics[width=1\columnwidth]{{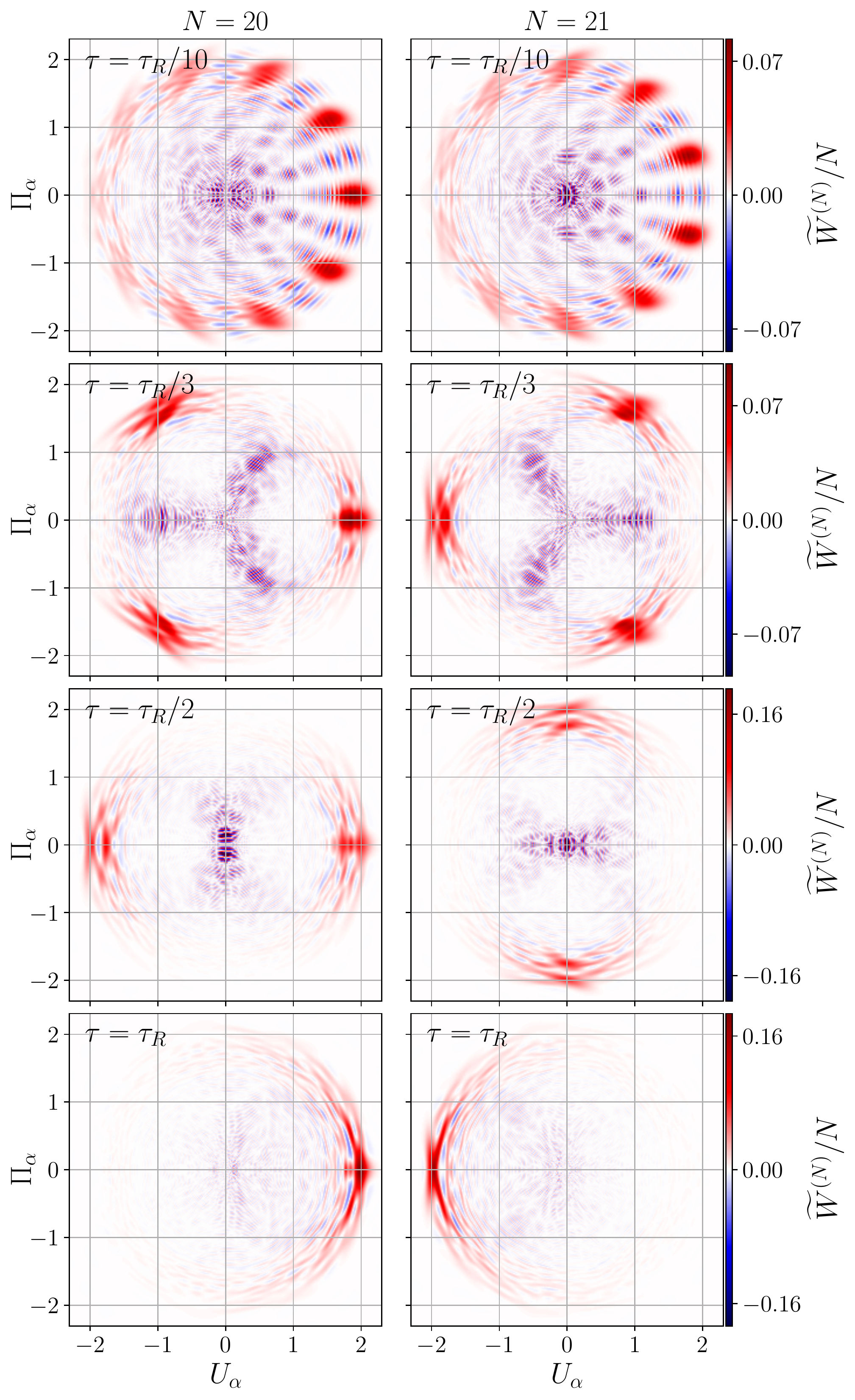}}
 \caption{Wigner functions $\wigner{\widetilde{W}}{N}{U_\alpha, \Pi_\alpha, t} / N$ for the initial value $\alpha_0 = 2$ for $N=20$ (left column) and $N=21$ (right column) at different times given in units of the revival time $\tau_R$.}
 \label{fig9}
\end{figure}
\begin{figure}
 \centering
 \includegraphics[width=1\columnwidth]{{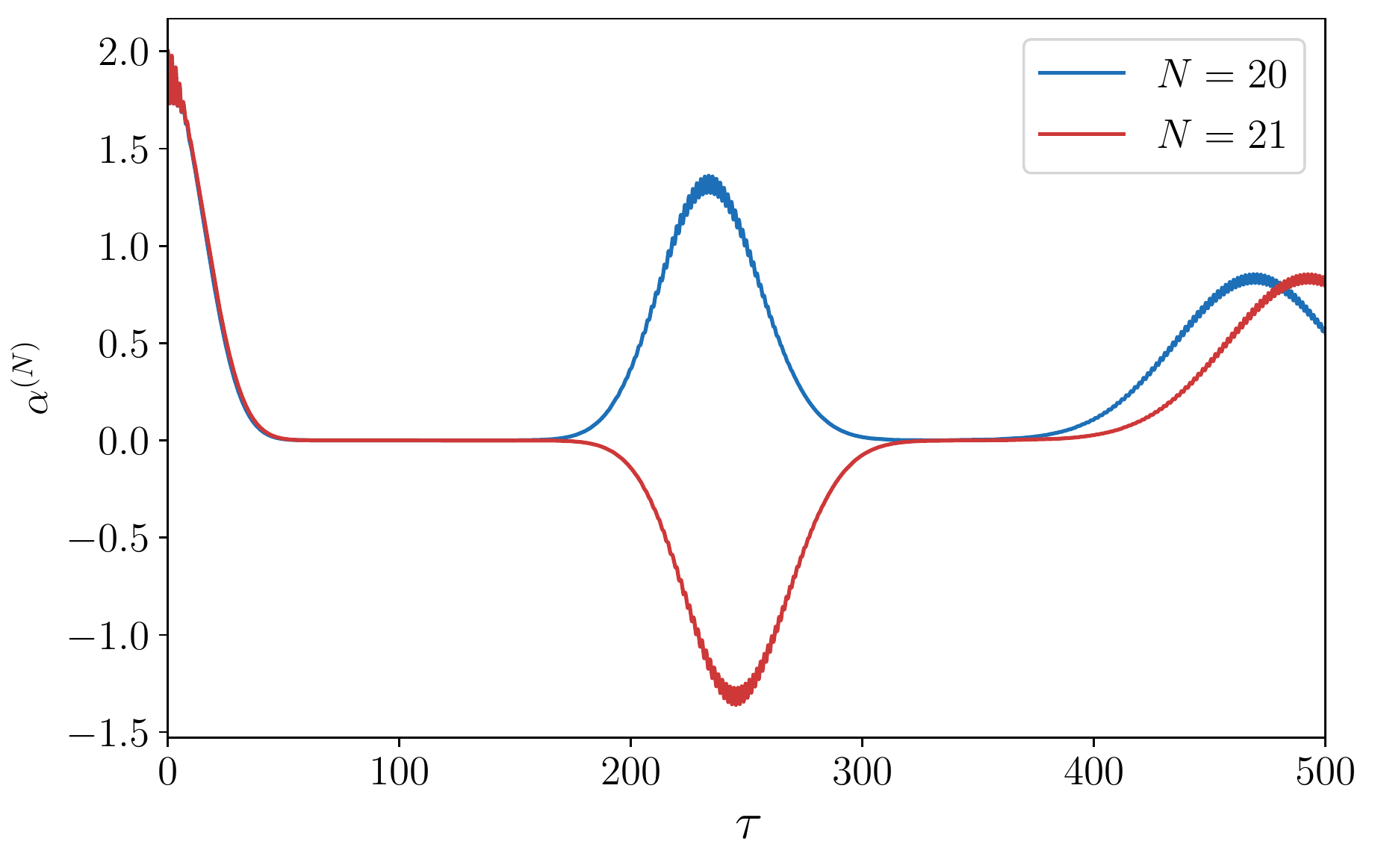}}
 \caption{Dynamics of $\highindex{\alpha}{N}$ for $\alpha_0 = 2$ and $N=20$ (blue) and $N=21$ (red) on a time scale including the first and second revival.}
 \label{fig10}
\end{figure}

In \cref{fig9} we have plotted the Wigner function for $N=20$ (left column) and $N=21$ (right column) QEs and for the initial condition $\alpha_0 = 2$ at four different times: $\tau=\tau_R/10$, $\tau=\tau_R/3$, $\tau=\tau_R/2$, and $\tau=\tau_R$ from top to bottom, where $\tau_R$ is the revival time extracted from our numerical calculations.
For a better interpretation of these results \cref{fig10} shows the temporal evolution of the coherent amplitude over a time scale including the first two revivals [similar to \cref{fig4}(d)].
At all times the overall structure of the Wigner function does not resemble a single Gaussian distribution anymore.
Instead, typically multiple localized structures (red areas) are seen with large interference-like patterns (red and blue stripes) in between.
Already at $\tau_R/10$ (upper panel in \cref{fig9}) the light field shows pronounced quantum features indicated by large negative contributions of the Wigner function.
When increasing the time from $\tau_R/10$ over $\tau_R/3$ to $\tau_R/2$ one observes that the number of localized (red) features reduces.
At $\tau_R/M$ there are exacly $M$ of those features visible until at $\tau_R$ (lower panel) only one single feature remains which is concentrated in a rather small region of the phase space.
There is however one striking difference between the case $N=20$ and $N=21$ despite the small change in the QE number.
At $\tau = \tau_R$ the Wigner function is strongly concentrated around $(U_\alpha, \Pi_\alpha) = (2,0)$ for $N=20$ and $(U_\alpha, \Pi_\alpha) = (-2,0)$ for $N=21$.
The resulting coherent amplitude in the frame rotating with the frequency of the light mode occurs with the same sign as the initial amplitude in the case of even $N$, while for odd $N$ it appears with opposite sign (see also \cref{fig10} and Ref.~\cite{jarvis2009dyn}).
This difference can be traced back to the fact that in the angular momentum representation [\cref{eq:H}] the TC model for even $N$ refers to a bosonic system while for odd $N$ it describes a fermionic system.

To understand the shape of the Wigner functions for different $\tau$ and $N$ it is instructive to refer to the limiting case of high mean photon numbers.
In this regime it has been shown, first for the JC model \cite{gea-banacloche1990col,gea-banacloche1991ato} and later for the TC model \cite{meunier2006ent,jarvis2009dyn}, that the state of the complete system can be written in the form
\begin{align}
\ket{\highindex{\Psi}{N}(t)}= \sum_{k=-N/2}^{N/2} \beta_k(t) \ket{D_k(t)} \otimes \ket{\Phi_k(t)}\ ,
\end{align}
where $\beta_k(t)$ are time-dependent coefficients, $\{\ket{D_k(t)}\}$ is a set of $N+1$ orthogonal states of the $N$ QEs defined, e.g., in Ref.~\cite{meunier2006ent}, and $\ket{\Phi_k(t)}$ are the so-called Gea-Banacloche states \cite{meunier2006ent,jarvis2009dyn}
\begin{align}\label{eq:gea}
 \ket{\Phi_k(t)}=\ket{\xi e^{i2\pi k t/t_R}}\ ,
 \end{align}
which are $N+1$ coherent states with mean photon number $|\xi|^2$.
In our Wigner functions defined in the scaled phase space $(U_\alpha, \Pi_\alpha)$ they correspond to $N+1$ Gaussians rotating with different frequencies on a circle with radius $\alpha_0$, located at time $\tau$ at $(U_\alpha, \Pi_\alpha)= (\alpha_0 \cos(2\pi k \tau/\tau_R), \alpha_0 \sin(2\pi k \tau/\tau_R))$.

At $\tau = 0$ all Gaussians overlap at $(U_\alpha, \Pi_\alpha) = (\alpha_0, 0)$. At the beginning of the dynamics the states with $k>0$ move in the positive $\Pi_\alpha$ direction, those with $k<0$ move in the negative $\Pi_\alpha$ direction.
States with different $|k|$ move with different velocities leading to the smearing out of the initial coherent state, as seen in \cref{fig6,fig7} (right column) and \cref{fig8} (c), where the width $\Delta_\pi^{(N)}$ increases much faster than $\Delta_u^{(N)}$.
As discussed in Ref.~\cite{meunier2006ent} this separation of the Gaussian distributions is the reason for the collapse of the Rabi oscillations.
As seen in \cref{fig4} (d) the collapse also leads to the deviation between the quantum and semiclassical model.

From Eq.~\eqref{eq:gea} we deduce that at the revival time $\tau=\tau_R$ all Gaussians overlap again, for even numbers $N$ (i.e., in the bosonic case) at $(U_\alpha, \Pi_\alpha) = (\alpha_0, 0)$ and for odd $N$ (i.e., in the fermionic case) at $(U_\alpha, \Pi_\alpha) = (-\alpha_0, 0)$ (see also Ref.~\cite{jarvis2009dyn}).
This is indeed what we see in the bottom panels of \cref{fig9}, except for the fact that the Wigner function, although very localized compared to the other depicted times, is significantly smeared out compared to a coherent state.
This is related to the fact that in our case with on average four photons per QE, we are still far from the limit of large photon numbers.

From Eq.~\eqref{eq:gea} we also find that at the times $\tau=\tau_R/M$ with $1 \leq M \leq N$ some of the Gaussians overlap, such that we observe $M$ isolated localizations of the Wigner function located at equal distances along the circle.
This is clearly seen in \cref{fig9} for the values $M=10, 3, 2, 1$ (from top to bottom).
These partial overlaps lead to small revivals of the Rabi oscillations at intermediate times \cite{meunier2006ent,jarvis2009dyn}, which are not visible in the photon amplitude $\alpha^{(N)}$ [\cref{fig4}], but very weakly in the inversion and the photon number [see \cref{fig5} (c) at $\tau \approx 120$ (blue curve), $\tau \approx 240$ (red curve) and $\tau \approx 360$ (green curve)].
Interestingly, between the maxima of the Wigner function there are pronounced negative contributions, which is a clear indication of non-classical behavior.
In particular, for $M=2$ the Wigner function has the qualitative shape of a Schr\"odinger cat state consisting of two maxima and an interference pattern in between.
This holds despite the fact that according to the general theory we do not expect a factorization into QE and field states \cite{jarvis2009dyn}, except for the case $N=1$, i.e., the JC model \cite{gea-banacloche1990col,gea-banacloche1991ato}.
Also for higher values of $M$, here shown for $M=3$ and $M=10$, we observe increasingly complex interference patterns between the maxima indicating coherences between the corresponding quasi-coherent states.

\section{Impact of dissipation}\label{sec.dissipation}
\noindent
So far we only discussed the pure Hamiltonian dynamics of the system.
In real systems the interaction of the system with external baths plays an important role for the internal dynamics of the quantum state and leads to dissipative effects.
JC \cite{gea-banacloche1993jay,puri1986col} and TC \cite{meunier2006ent,munoz2019sym,dhar2018var,lemini2018bou} models with dissipation, often in the presence of an external driving, have been widely studied.
In this section we study the impact of dissipation on the temporal and spectral properties as well as on the Wigner functions for our system.
We assume that the lifetime of the TLS is much longer than the lifetime of photons in the resonator mode, such that we only take into account the decay of the photon mode (e.g., due to photon losses through the mirrors of the cavity) with a rate $\Gamma$ \cite{meunier2006ent,munoz2019sym}.

To analyze the changes induced by the damping we solve the Master equation in Lindblad form with the dissipation rate $\Gamma$, which in the frame scaled with the dimensionless dissipation rate $\widetilde{\Gamma} = \Gamma N / (G \sqrt{N})$ is given by
\begin{align}
   \dv{\tau} \hat{\rho}^{(N)} &= -\frac{i}{\hbar G \sqrt{N}} \left[ \hat{H}, \hat{\rho}^{(N)} \right] \nonumber \\
                    &+ \widetilde{\Gamma} \left[ \hat{\alpha} \hat{\rho}^{(N)} \hat{\alpha}^\dagger  - \frac{1}{2} \left( \hat{\rho}^{(N)} \hat{\alpha}^\dagger \hat{\alpha} + \hat{\alpha}^\dagger \hat{\alpha} \hat{\rho}^{(N)} \right)\right],
\end{align}
use the same initial condition as above
\begin{align}
   \hat{\rho}^{(N)}(0) &= \ket{\Psi^{(N)}(0)} \bra{\Psi^{(N)}(0)}, \nonumber
\end{align}
calculate the expectation value
\begin{align}
    \alpha^{(N)}(\tau) = \Tr[\hat{\rho}^{(N)}(\tau) \hat{\alpha}],\nonumber
\end{align}
and the corresponding Wigner functions in the scaled frame $\wigner{\widetilde{W}}{N}{U_\alpha, \Pi_\alpha, \tau}$.
\begin{figure}
    \centering
    \includegraphics[width=1\columnwidth]{{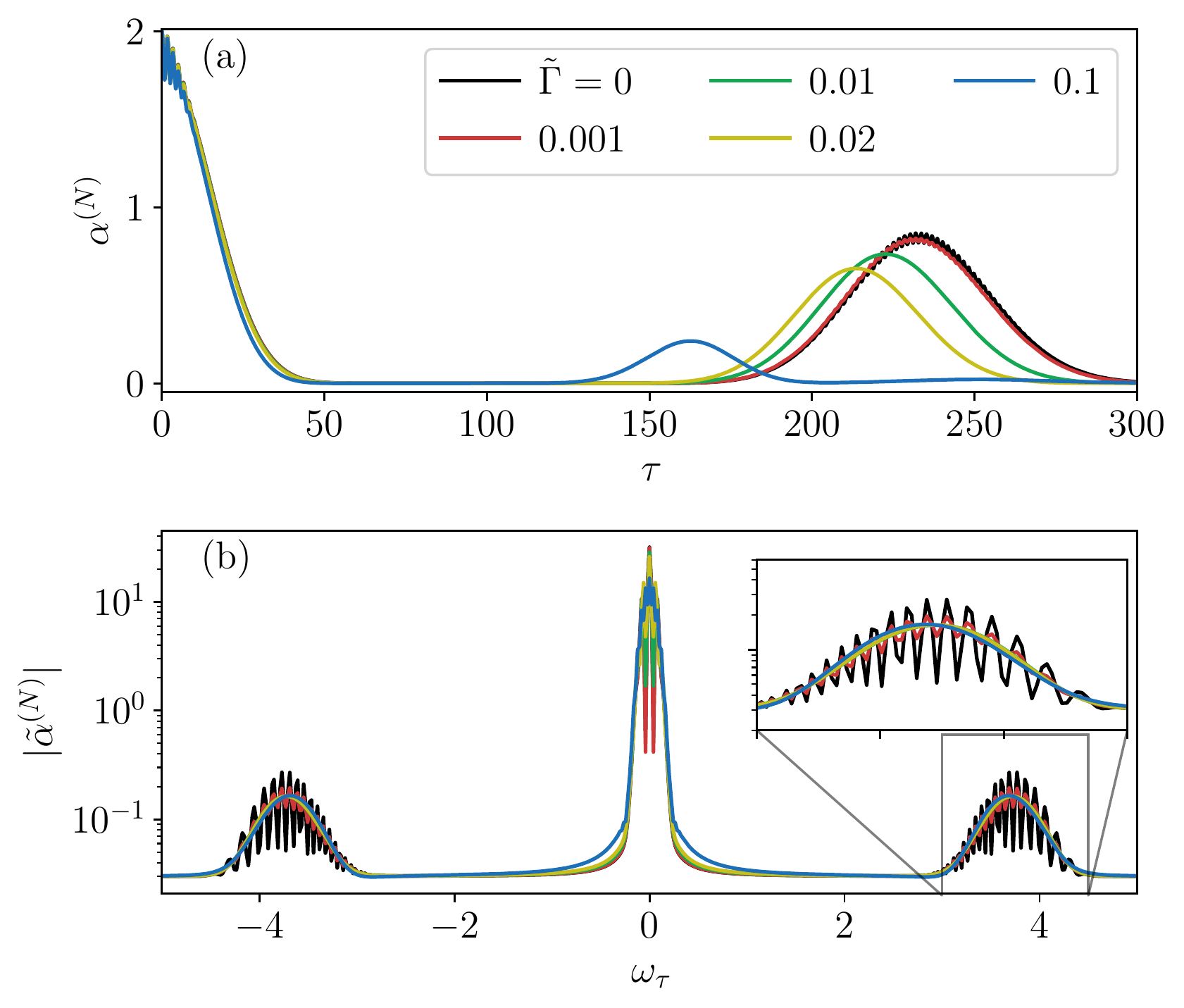}}
    \caption{(a) Dynamics of $\highindex{\alpha}{N}$ for $N=20$ with various damping rates $\widetilde{\Gamma}$ and (b) corresponding spectra. The inset shows an enlargement of the region of the right side peaks.}
    \label{fig11}
\end{figure}

Figure~\ref{fig11} shows the dynamics of the coherent amplitude $\highindex{\alpha}{N}(\tau)$ in (a) and the corresponding spectra in (b) for various damping rates $\widetilde{\Gamma}$ and $N=20$.
In \cref{fig11} (a) we clearly see that the damping has a remarkable influence on the strength and the time of the revival.
The revival time reduces for increasing $\widetilde{\Gamma}$ (as was similarly observed in the damped JC model \cite{puri1986col}), which is not surprising since the revival time depends on the average number of photons and the dissipation is associated with the loss of photons. 
We also observe that the amplitude of the weak oscillations during the revival is significantly reduced even for relatively small dissipation.
As seen in the spectra in \cref{fig11} (b), this damping leads to a broadening of the individual lines constituting the side peaks, such that already for $\widetilde{\Gamma} \approx 0.01$ smooth side peaks emerge, while in the central peak the individual lines are still visible.
If the damping increases further and reaches values larger than the inverse collapse time, the revival vanishes and the initial decay of the  coherent amplitude becomes faster.
This leads to a strong broadening also of the central line in the spectrum (not shown).
\begin{figure}
    \centering
    \includegraphics[width=1\columnwidth]{{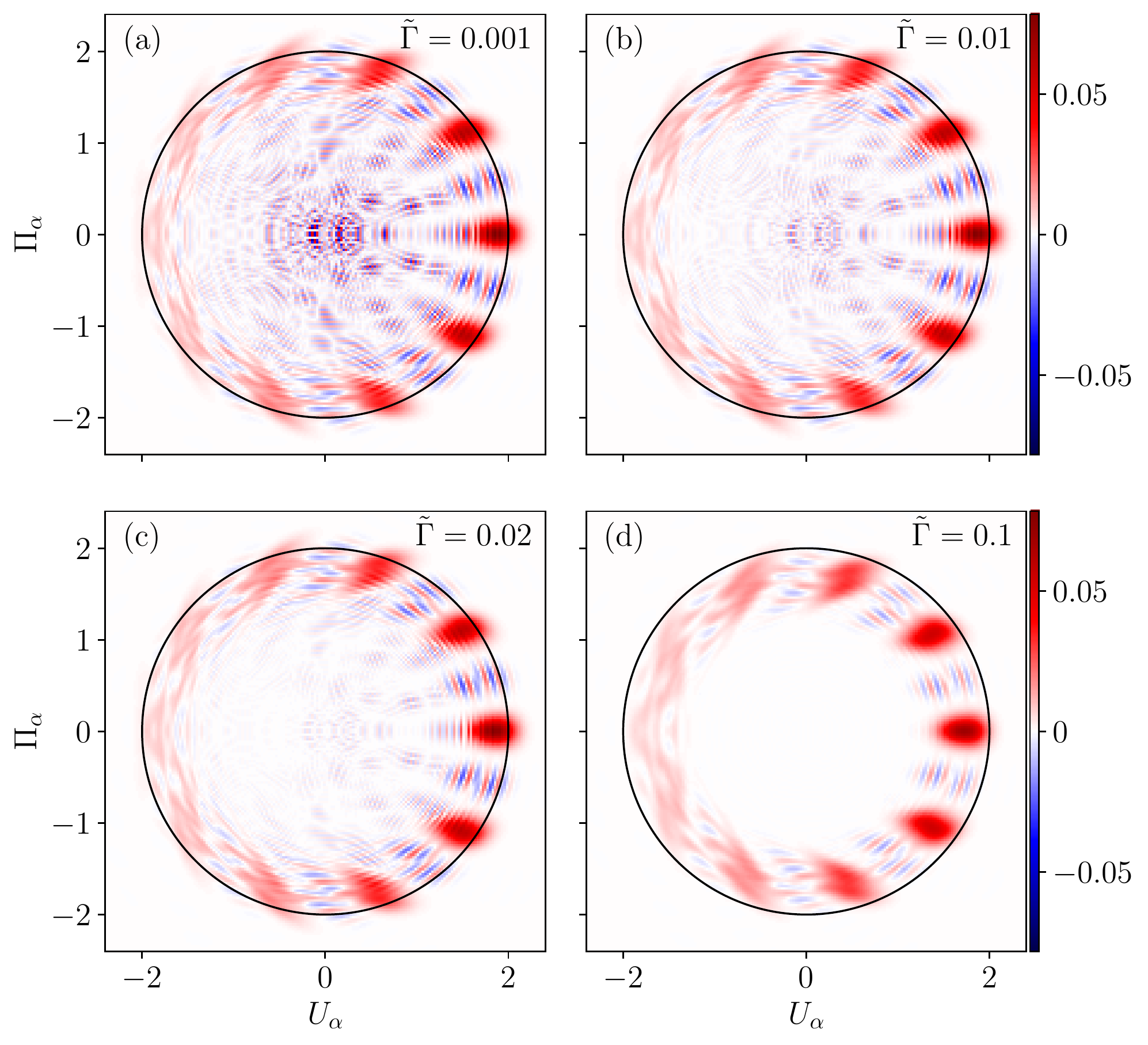}}
    \caption{Transformed Wigner functions $\wigner{\widetilde{W}}{N}{U_\alpha, \Pi_\alpha, \tau} / N$ at the time $\tau_R / 10$ for different damping rates $\widetilde{\Gamma}$. The value of $\tau_R$ refers to the revival time of the undamped system. The black circles correspond to the trajectory on which the initial Gaussian would move for $\widetilde{\Gamma} = 0$.}
    \label{fig12}
\end{figure}

In \cref{fig12} the Wigner functions $\wigner{\widetilde{W}}{N}{U_\alpha, \Pi_\alpha, \tau_R/10}$ at the time $\tau = \tau_R/10$ are plotted for different damping rates $\widetilde{\Gamma}$, where $\tau_R$ refers to the revival time of the undamped system. The black circle marks the trajectory of the initial Gaussian without any coupling or dissipation.
By comparing the position of the peaks and the black circle one can see, that the entire distribution shrinks towards the center of the phase space reflecting the loss of photons from the system.
For increasing $\widetilde{\Gamma}$ we clearly see that the interferences in the middle of the circle vanish first, while the ones on the circumference of the circle between neighboring peaks remain present for relatively large $\widetilde{\Gamma} > 0.1$. Indeed, as has been shown in Ref.~\cite{meunier2006ent}, the decoherence functional depends on the phase difference $\Delta \phi_{k, l} = \phi_k - \phi_l$ between the Gea-Banacloche states $\ket{\Phi_k(\tau)} = \ket{\xi e^{i\phi_k}}$ with $\phi_k = 2\pi k \tau /\tau_R$ when neglecting the loss of photons.
Due to the dissipation large values of $\Delta \phi_{k, l}$ are averaged out on shorter time scales than small $\Delta \phi_{k, l}$.
In \cref{fig12} the coherences on the circumference are induced by neighboring peaks (small $\Delta \phi_{k, l}$) and the ones in the center by opposing lines (large $\Delta \phi_{k, l}$), which explains why the coherences in the center are damped away earlier than the ones on the circumference.

\section{Conclusions}\label{sec.conclusions}
We have presented a detailed comparison of the spectral and temporal characteristics obtained from a semiclassical and a fully quantum optical treatment of an ensemble of QEs in a cavity, in which a coherent initial state is created by excitation with an ultrashort laser pulse.
We have found that the spectra of the quantum system show the same features as the semiclassical system even for relatively small numbers of QEs, however, they exhibit a broadening which decreases with increasing number of QEs.
The collective width seen in the spectra can be traced back to the finite collapse time observed in the dynamics of the quantum system.
Because of this collapse, which is absent in the semiclassical case, the coherent field amplitude matches only on small time scales.
During this collapse time, however, the frequencies agree very well leading to the similarity of the spectra.
Adding the effect of dissipation, we explicitly showed that with increasing damping rate the side peaks in the Rabi regime and for higher rates also the central peak are broadened until eventually the revival vanishes.

By investigating the temporal evolution of the Wigner function for the cavity mode at small times, it was possible to show that a good agreement in the expectation values is in general not associated with a quasi-classical behavior of the quantum state.
In the Wigner functions we found squeezing at short times and strong broadenings, in particular in the $\Pi_\alpha$ direction, on time scales where the temporal field characteristics of the semiclassical and the quantum model still agree very well.
The deviations from a classical state thus seem to have only a minor influence on the expectation values on these time scales.
Nonetheless we have found that the semiclassical model describes the complete quantum system very well for small initial field values, i.e., small amplitudes of the exciting laser pulse.

We then have studied the Wigner function at different times during the collapse and revival of the coherent field amplitude.
At the revival time $\tau_R$ the Wigner function is indeed concentrated in the phase space region of a corresponding coherent state, the shapes, however, for the values of the mean photon numbers studied here deviate strongly from the circular shape, expected for a coherent state.
At the times $\tau_R/M$ for $1 \leq M \leq N$ the Wigner function exhibits the structure of Schr\"odinger cat states consisting of a superposition of $M$ quasi-coherent states distributed on a circle around the origin in phase space.
By additionally accounting for dissipation, we showed that for increasing dissipation rates the coherences in the center of the circle are destroyed first, while the ones between neighboring states are much more robust.

\section*{Acknowledgments}
DW thanks NAWA for financial support within the ULAM program (No. PPN/ULM/2019/00064).

\appendix
\section{Calculation of the Wigner function}\label{app:wigner}
Here we calculate the Wigner function of the system in the original, non-rotating frame \cite{haroche2006exp}.
In the basis of eigenstates $\ket{j, n; l}$ the Wigner function reads
\begin{align}
  \wigner{W}{N}{\lambda, t} = \sum_{\substack{n, n' \\l, l'}} \rho^{(N, n, n')}_{l, l'}(t) w^{(N, n, n')}_{l, l'}(\lambda) \nonumber
\end{align}
with 
\begin{align}
w^{(N, n, n')}_{l, l'}(\lambda) &= \frac{2}{\pi} \Tr[D(-\lambda) \kb{j, n; l}{j, n'; l'} D(\lambda) \mathcal{P}], \nonumber \\
\rho^{(N, n, n')}_{l, l'}(t) &= \exp(-\abs{\xi}^2) \frac{\xi^n}{\sqrt{n!}} \frac{\xi^{*n'}}{\sqrt{n'!}} \ccoeff{c}{0, l}{N,n} \coeff{c}{0, l'}{N,n'} \nonumber \\
&\times \exp(-\frac{i}{\hbar} \left( E_l^{(N, n)} - E_{l'}^{(N, n')} \right) t ). \nonumber
\end{align}
The Wigner functions in the basis of the eigenstates can be expressed by the Wigner functions in the Fock basis
\begin{align}
 w^{(N, n, n')}_{l, l'}(\lambda) &= \frac{2}{\pi} \sum_{k=0}^{\min(N, n)} \coeff{c}{k, l}{N, n} \ccoeff{c}{k, l'}{N, n'} \nonumber \\
    &\times \Tr[D(-\lambda) \kb{n-k}{n'-k} D(\lambda) \mathcal{P}] \nonumber \\
    &= \sum_{k=0}^{\min(N, n)} \coeff{c}{k, l}{N, n} \ccoeff{c}{k, l'}{N, n'} w_F^{(n-k,n'-k)}(\lambda),\nonumber
\end{align}
where the property $\braket{j, -j+k}{j, -j+k'} = \delta_{k, k'}$ and the cyclic invariance under the trace have been used.
In the Fock basis the Wigner function reads
\begin{align}
 w_F^{(n,n')}(\lambda) &= \frac{2}{\pi} \Tr[D(-\lambda) \kb{n}{n'} D(\lambda) \mathcal{P}] \nonumber \\
    &= \frac{2}{\pi} \Tr[ \mathcal{P} \kb{n}{n'} D(\lambda) \mathcal{P} D(-\lambda)  \mathcal{P}^\dagger] \nonumber \\
    &= \frac{2}{\pi} \Tr[ \mathcal{P} \kb{n}{n'} D(\lambda) D(\lambda) ] \nonumber \\
    &= \frac{2}{\pi} \left( -1 \right)^{n} \Tr[\kb{n}{n'} D(2\lambda)] \nonumber \\
    &= \frac{2}{\pi} \left( -1 \right)^{n} C_s^{(n,n')} (2\lambda) \nonumber
\end{align}
with the symmetric characteristic function $C_s^{(n,n')} (x)$, which is for $n' \geq n$ given by \cite{cahill1969ord}
\begin{align}
 C_s^{(n,n')} (x) &=  \exp(-\frac{\abs{x}^2}{2}) x^{n'-n} \nonumber \\
    &\times \sqrt{\frac{n!}{n'!}} L_{n}^{(n'-n)}(\abs{x}^2), \nonumber \\
 L_{m}^{(n)}(x) &= \sum_{j=0}^{m} (-1)^j \frac{x^j}{j!} 
    \begin{pmatrix}m + n \\ m-j\end{pmatrix}, \nonumber
\end{align}
where $L_{n}^{(n'-n)}(x)$ are the generalized Laguerre polynomials. For $n' < n$ the characteristic function is obtained
using the following symmetry property:
\begin{align}
 C_s^{(n',n)} (x) = C_s^{(n,n')*}(-x). \nonumber
\end{align}

\section{Transformation of the Wigner function}\label{app:wigner_transformation}
The observables of the light field can be expressed in terms of the Wigner function using the quadrature operators $\hat{u} = \frac{1}{2} \left( \hat{a} + \hat{a}^\dagger \right)$ and $\hat{\pi} = \frac{1}{2i} \left( \hat{a} - \hat{a}^\dagger \right)$, e.g.,
\begin{align}
 \erw{\hat{a}} &= \erw{\hat{u}} + i \erw{\hat{\pi}} \nonumber \\
    &= \int \dd[2]{\lambda} \lambda \wigner{W}{N}{\lambda, t} , \nonumber \\
 \erw{\hat{a}^\dagger \hat{a}} &= \erw{\hat{u}^2} + \erw{\hat{\pi}^2} - \frac{1}{2} \nonumber \\
 &= \int \dd[2]{\lambda} \left( U^2 + \Pi^2 \right) \wigner{W}{N}{\lambda, t} - \frac{1}{2} \nonumber 
\end{align}
with $\lambda = U + i\Pi$.
We now use the transformation from \cref{eq:trafo} to relate the obervables in the rotated frame to the Wigner function.
The transformed field amplitude is given by
\begin{align}
 \erw{\hat{\alpha}}^{(N)} &= \frac{e^{i\omega_0 t}}{\sqrt{N}} \int \dd[2]{\lambda} \lambda \wigner{W}{N}{\lambda, t} \nonumber \\
    &= \frac{1}{\sqrt{N}} \int \dd{\abs{\lambda}} \dd{\phi} \abs{\lambda}^2 e^{i(\phi + \omega_0 t)} \wigner{W}{N}{\abs{\lambda}e^{i\phi}, t} \nonumber \\
    &= \frac{1}{\sqrt{N}} \int \dd{\abs{\lambda}} \dd{\phi'} \abs{\lambda}^2 e^{i\phi'} \wigner{W}{N}{\abs{\lambda} e^{i\phi'} e^{-i\omega_0 t}, t} \nonumber \\
    &= \frac{1}{\sqrt{N}} \int \dd{U} \dd{\Pi} \left(U + i\Pi\right) \highindex{W}{N}\left(\left( U + i\Pi \right)e^{-i\omega_0 t}, t \right), \nonumber
\end{align}
where we used the $2\pi$ periodicity of the integrand in polar coordinates.
Further introducing the quadrature coordinates in the transformed frame $U_\alpha = U / \sqrt{N}$ and $\Pi_\alpha = \Pi / \sqrt{N}$ the transformed field amplitude reads
\begin{align}
 \erw{\hat{\alpha}}^{(N)} &= \int \dd{U_\alpha} \dd{\Pi_\alpha} \left(U_\alpha + i\Pi_\alpha \right) \nonumber \\
    &\times N \highindex{W}{N}\left(\sqrt{N}\left( U_\alpha + i\Pi \right)e^{-i\omega_0 t}, t \right) \nonumber \\
    &= \int \dd{U_\alpha} \dd{\Pi_\alpha} \left(U_\alpha + i\Pi_\alpha \right) \wigner{\widetilde{W}}{N}{U_\alpha, \Pi_\alpha, \tau} \nonumber \\
    &= \erw{\hat{u}_\alpha}^{(N)} + i \erw{\hat{\pi}_\alpha}^{(N)} \nonumber
\end{align}
with the transformed quadrature operators $\hat{u}_\alpha = \hat{u} / \sqrt{N} = \frac{1}{2} \left(\hat{\alpha} + \hat{\alpha}^\dagger \right)$, $\hat{\pi}_\alpha = \hat{\pi} / \sqrt{N} = \frac{1}{2i} \left(\hat{\alpha} - \hat{\alpha}^\dagger \right)$ and the transformed Wigner function at the dimensionless time $\tau$
\begin{align}
&\wigner{\widetilde{W}}{N}{U_\alpha, \Pi_\alpha, \tau} \nonumber \\
&\quad = N \, \wigner{W}{N}{\sqrt{N} (U_\alpha + i\Pi_\alpha) e^{-i\omega_0 \frac{\tau}{g\sqrt{N}}}, \frac{\tau}{g\sqrt{N}}}. \nonumber 
\end{align}


%

\end{document}